\documentclass[prd,aps,twocolumn,nofootinbib,showpacs]{revtex4-1}

\usepackage{amsfonts}
\usepackage{amsmath}
\usepackage{amssymb}
\usepackage{bm}
\usepackage{dcolumn}
\usepackage[dvips]{graphicx}
\usepackage{graphics}
\usepackage[latin1]{inputenc}
\usepackage{latexsym}
\usepackage{rotating}
\usepackage[colorlinks=true]{hyperref}
\usepackage{xspace} 
\usepackage[usenames]{color}
\usepackage{mathrsfs}
\usepackage{multirow}
\usepackage{pifont}
\usepackage{enumitem}

\definecolor {darkgreen}{rgb}{0.2,0.7,0.2}
\definecolor{purple}{rgb}{0.5,0,0.5}

\newcommand{\eq}{\begin{equation}}
\newcommand{\be}{\begin{equation}}
\newcommand{\eeq}{\end{equation}}
\newcommand{\ee}{\end{equation}}

\newcommand{\GW}{{\mbox{\tiny GW}}}
\newcommand{\ppE}{{\mbox{\tiny ppE}}}
\newcommand{\IM}{{\mbox{\tiny IM}}}

\newcommand{\MR}{{\mbox{\tiny MR}}}

\newcommand{\GR}{{\mbox{\tiny GR}}}
\newcommand{\dip}{{\mbox{\tiny Dip}}}
\newcommand{\ED}{{\mbox{\tiny ED}}}
\newcommand{\EA}{{\mbox{\tiny EA}}}
\newcommand{\KG}{{\mbox{\tiny KG}}}
\newcommand{\MG}{{\mbox{\tiny MG}}}
\newcommand{\MDR}{{\mbox{\tiny MDR}}}

\newcommand{\low}{{\mbox{\tiny low}}}
\newcommand{\hi}{{\mbox{\tiny high}}}
\newcommand{\hicut}{{\mbox{\tiny high-cut}}}
\newcommand{\locut}{{\mbox{\tiny low-cut}}}
\newcommand{\ratiolo}{{\mbox{\tiny lratio}}}
\newcommand{\ratiohi}{{\mbox{\tiny hratio}}}
\newcommand{\isco}{{\mbox{\tiny isco}}}
\newcommand{\yrs}{{\mbox{\tiny 3 years}}}
\newcommand{\spc}{{\mbox{\tiny space}}}
\newcommand{\grnd}{{\mbox{\tiny ground}}}

\newcommand{\sd}{\partial}
\newcommand{\tilh}{\tilde{h}}
\newcommand{\Ln}{\mbox{ln}}
\newcommand{\scM}{\mathcal{M}}
\newcommand{\scA}{\mathcal{A}}

\usepackage{tabularx}
\usepackage{array}
\usepackage{tabulary}
\renewcommand{\arraystretch}{1.5}
\newcolumntype{Y}{>{\centering\arraybackslash}X}
\newcolumntype{S}{>{\hsize=.2\hsize}X}
\newcolumntype{M}{>{\centering\arraybackslash}S}

\begin{document}
\title{Theoretical Physics Implications of \\ Gravitational Wave Observation with Future Detectors}

\author{Katie Chamberlain}
\author{Nicol\'as Yunes}
\affiliation{eXtreme Gravity Institute, Department of Physics, Montana State University, Bozeman, MT 59717, USA.}

\begin{abstract}
Gravitational waves encode invaluable information about the nature of the relatively unexplored \emph{extreme gravity} regime, where the gravitational interaction is strong, non-linear and highly dynamical. 
Recent gravitational wave observations by advanced LIGO have provided the first glimpses into this regime, allowing for the extraction of new inferences on different aspects of theoretical physics.
For example, these detections provide constraints on the mass of the graviton, Lorentz violation in the gravitational sector, the existence of large extra dimensions, the temporal variability of Newton's gravitational constant, and modified dispersion relations of gravitational waves. 
Many of these constraints, however, are not yet competitive with constraints obtained, for example, through Solar System observations or binary pulsar observations.
In this paper, we study the degree to which theoretical physics inferences drawn from gravitational wave observations will strengthen with detections from future detectors.
We consider future ground-based detectors, such as the LIGO-class expansions A+, Voyager, Cosmic Explorer and the Einstein Telescope, as well as space-based detectors, such as various configurations of eLISA and the recently proposed LISA mission.
We find that space-based detectors will place constraints on General Relativity up to 12 orders of magnitude more stringently than current aLIGO bounds, but these space-based constraints are comparable to those obtained with the ground-based Cosmic Explorer or the Einstein Telescope (A+ and Voyager only lead to modest improvements in constraints). 
We also generically find that improvements in the instrument sensitivity band at low frequencies lead to large improvements in certain classes of constraints, while sensitivity improvements at high frequencies lead to more modest gains.
These results strengthen the case for the development of future detectors, while providing additional information that could be useful in future design decisions.

\end{abstract}

\date{\today}
\maketitle

\section{Introduction}
The recent detection of gravitational waves by the advanced LIGO (aLIGO) detectors~\cite{PhysRevLett.116.061102,Abbott:2016nmj} has revealed that the gravitational interaction seems to behave as predicted by Einstein's theory of General Relativity~(GR), even in the \emph{extreme gravity regime}~\cite{TheLIGOScientific:2016src} where the gravitational interaction is strong, non-linear and highly dynamical~\cite{Yunes:2013dva}. This confirmation of GR in extreme gravity can then be used to extract inferences on fundamental theoretical physics by imposing generic constraints on deviations from the pillars of Einstein's theory, i.e.~constraints on physical mechanisms that may or may not activate in the extreme gravity regime~\cite{Yunes:2016jcc}. Examples of these inferences include statements about the mass of the graviton and thus about its propagation speed (as predicted e.g.~in new massive gravity~\cite{Rubakov:2004eb,Hinterbichler:2011tt,deRham:2014zqa} and bigravity~\cite{Hassan:2011zd}), whether Lorentz invariance is violated in the gravitational sector (as predicted e.g.~in Einstein-\AE{}ther theory~\cite{Jacobson:2008aj,Jacobson:2000xp} and khronometric gravity~\cite{Blas:2009ck,Blas:2010hb}), the existence of a large extra dimension (as predicted e.g.~in certain Randall-Sundrum scenarios~\cite{Randall:1999ee,Randall:1999vf}), the temporal variability of fundamental physical constants~\cite{Yunes:2009bv}, and the dispersion relation of gravitational waves (modifications of which are predicted in a plethora of quantum gravitational models~\cite{AmelinoCamelia:2000ge,Magueijo:2001cr,Kostelecky:2016kfm,Calcagni:2009kc,Calcagni:2011kn,Calcagni:2011sz}). 

Most of the constraints one can place on GR deviations with gravitational waves, however, are not yet competitive with constraints derived with Solar System or binary pulsar observations~\cite{Will:2014kxa}. This is in part because of the much higher signal-to-noise ratio of Solar System and binary pulsar observations relative to current gravitational wave observations. But this will undoubtably change as next-generation gravitational wave observatories begin operation in the coming decades with much higher signal-to-noise ratio observations. Although the future is uncertain, there are proposals to upgrade the aLIGO instrument into an A+ and a Voyager configuration that would become operational by $\sim 2020$ and $\sim 2027$ respectively~\cite{LIGO-Tech-doc}. There are also plans to construct an entirely new next-generation observatory, either Cosmic Explorer or Einstein Telescope, by the middle of the following decade~\cite{LIGO-Tech-doc}. We will describe some of the details of these upgrades later, but suffice it to say that they will allow for observations with signal-to-noise ratios that are 6 times, 12 times, and 60 times larger than current aLIGO. By the mid 2030s, space-based gravitational wave observatories, such as the proposed LISA~\cite{Audley:2017drz}, should also begin operation, allowing us to observe gravitational waves emitted by much more massive compact objects at very high signal-to-noise ratios.

This paper addresses the degree to which inferences on theoretical physics will become stronger given gravitational wave observations with future detectors. We have already shown that inferences on the existence of dipole radiation in the merger of compact binaries can be strengthened by 5--6 orders of magnitude with joint aLIGO-evolved LISA observations~\cite{Barausse:2016eii}. We now extend this analysis to consider (i) several other theoretical physics mechanisms that can be constrained\footnote{Indeed, generic theoretical physics mechanisms can be constrained with gravitational waves in a model-independent way, such as Lorentz invariance or parity violation, as explained in detail in~\cite{Yunes:2016jcc}.} and (ii) several other future detectors (in particular, upgrades to ground-based detectors) that will allow for much louder future observations. We carry out a large number of Fisher analysis calculations assuming single events, sky averaging, and use both realistic and phenomenological sensitivity curves. The latter are included when developing a physical and mathematical understanding of the results obtained using realistic sensitivity curves. These Fisher calculations allow us to estimate projected constraints on a variety of physical mechanisms, and also on particular modified theories of gravity, as a function of the post-Newtonian (PN)\footnote{In the PN approximation, one solves the field equations as an expansion in low velocities (relative to the speed of light) and weak gravitational fields. For details/discussion, see~\cite{Blanchet:2006zz}.} order at which the modifications first enter the gravitational wave observable. 

The results obtained in this paper strengthen the fundamental science case for future gravitational wave detectors, both ground and space-based, and they provide information that could be used in design decisions as these new detectors are developed. For this reason, given the length of the paper and the fact that the results obtained may be interesting to different communities, we provide a short summary of our main findings below.
{\newcommand{\minitab}[2][l]{\begin{tabular}{#1}#2\end{tabular}}
\renewcommand{\arraystretch}{1.4}
\begingroup
\squeezetable
\begin{table*}[htb]
\begin{centering}
\begin{tabular}{c|c|c|c|c|c|c|c}
\hline
\hline
\noalign{\smallskip}
 GR Deviation &  PN & Parameter & Best Space Const. & Best Ground Const. & Current Const. & Best Space Sys. & Best Ground Sys. \\ 
\hline \hline
 \multirow{2}{*}{Dipole Radiation}  & \multirow{2}{*}{-1} & 
 $\beta$ & $4.9 \times 10^{-12}$  & $1.9\times 10^{-10}$ & $4.4\times 10^{-5}$ &  EMRI & NSNS \\
 & & $\delta \dot{E}_{\dip}$  & $7.8\times 10^{-8}$  & $3.2\times 10^{-8}$ & $1.8\times 10^{-3}$ &  EMRI/GW150914 & NSNS \\ \hline 
  \multirow{2}{*}{Large Extra-Dimension} & \multirow{2}{*}{-4} & 
 $\beta$ & $2.2\times 10^{-22}$  & $6.4\times 10^{-20}$ & $9.1\times 10^{-11}$ &  EMRI & NSNS \\
 & & $\ell$\ [$\mu m$]  & $3.0\times 10^{2}$  & $7.5\times 10^{4}$ & $10-10^{3}$\ \cite{Johannsen:2008tm,Johannsen:2008aa,Adelberger:2006dh,Psaltis:2006de,Gnedin:2009yt} &  EMRI/GW150914 & BHBH \\ \hline 
  \multirow{2}{*}{Time-Varying $G$}  & \multirow{2}{*}{-4} & 
 $\beta$ & $2.2\times 10^{-22}$  & $6.4\times 10^{-20}$ & $9.1\times 10^{-11}$ &  EMRI & NSNS \\
 & & $\dot{G}\ [1/yr]$  & $6.8\times 10^{-8}$  & $1.1\times 10^{-3}$ & $10^{-12}-10^{-13}$\ \cite{Bambi:2005fi,Copi:2003xd,Manchester:2015mda,2011Icar..211..401K,Hofmann} &  EMRI & NSNS \\ \hline 
  \multirow{2}{*}{Einstein-\AE{}ther Theory}  & \multirow{2}{*}{0} & 
 $\beta$ & $4.0\times 10^{-8}$  & $6.7\times 10^{-5}$ & $3.4\times 10^{-3}$ &  EMRI & $\ell$BHNS \\
 & & $(c_{+},c_{-})$  & $(10^{-3},3\times10^{-4})$  & $(10^{-2},4\times 10^{-3})$ & $(0.03,0.003)$\ \cite{Yagi:2013qpa,Yagi:2013ava} &  EMRI & NSNS \\ \hline 
  \multirow{2}{*}{Khronometric Gravity}  & \multirow{2}{*}{0} & 
 $\beta$ & $4.0\times 10^{-8}$  & $6.7\times 10^{-5}$ & $3.4\times 10^{-3}$ &  EMRI & $\ell$BHNS \\
 & & $(\beta_{\KG},\lambda_{\KG})$  & $(10^{-4},10^{-2})/2$  & $(10^{-2},10^{-1})/5$ & $(10^{-2},10^{-1})/2$\ \cite{Yagi:2013qpa,Yagi:2013ava} &  EMRI & GW150914 \\ \hline 
  \multirow{2}{*}{Graviton Mass}  & \multirow{2}{*}{+1} & 
 $\beta$ & $4.3\times 10^{-5}$  & $1.0\times 10^{-3}$ & $8.9\times 10^{-2}$ &  EMRI/IMBH & $\ell$BHBH \\
 & & $m_{g}\ [eV]$  & $9.0\times 10^{-28}$  & $9.9\times 10^{-25}$ & $10^{-29}- 10^{-18}$\ \cite{Finn:2001qi,Brito:2013wya,Hare:1973px,PhysRevD.9.1119,PhysRevLett.61.1159} &  SMBH/IMRI & GW150914 \\ \hline 
\noalign{\smallskip}
\hline
\hline
\end{tabular}
\end{centering}
\caption{Table summary of the best constraints on a variety of modified gravity modifications, listed in the first column. The second column indicates the PN order at which the modification first enters the gravitational wave phase. The third column labels the parameters that can be constrained. The fourth (fifth) column shows the best projected constraint achievable with a space-based (ground-based) detectors, which is to be compared with current constraints on $\beta$ (listed as the best constraint obtained with either of the GW150914 or GW151226 detections), and with current constraints on theory parameters as given by the most stringent of either aLIGO \textit{or} other observations. The last two columns show the class of the system that lead to the best constraint. Constraints on Einstein-\AE ther/khronometric Gravity are given as rough constraints on ($c_{+},c_{-}$)/($\beta_{\KG},\lambda_{\KG}$) (for the contours, see Figs.~\ref{fig:KG} and~\ref{fig:EA}).}
\label{tab:summary2}
\end{table*}
\endgroup}

\vspace{0.6cm}
\fbox{\parbox{8cm}{
{\bf Generic constraints with space-based detectors for deviations that enter first at negative (positive) PN order are constrained 12 (3) orders of magnitude better than current LIGO constraints, but comparable and at most 3 orders of magnitude better than constraints with third-generation ground detectors.}
}}
\vspace{0.0cm}

Constraints on GR deviations with space-based detectors are certainly much more stringent than those that can be placed with current ground-based detectors, if for no other reason than because the former will detect gravitational waves with a signal-to-noise ratio in the thousands. The constraints become comparable, however, when one considers what is achievable with third-generation detectors, since these will also observe in the very high signal-to-noise ratio regime.  The improvement of space-over-ground constraints is significantly larger when considering deviations that enter at negative PN order; this is simply because the latter have access to a much lower frequency band, allowing for the detection of low-mass binaries in the very early stages of inspiral.

\vspace{0.6cm}
\fbox{\parbox{8cm}{
{\bf  Improvements of ground-based detectors within the current LIGO facilities will lead to only modest improvements of constraints on GR deviations.}
}}
\vspace{0.0cm}

A+ and Voyager type improvements of the current LIGO facilities will lead to improved constraints on GR deviations that will not exceed an order of magnitude. This is because such projected modifications \textit{will not} greatly improve the low-frequency band of the detector noise (see Fig.~\ref{fig:Sncurves}). Cosmic Explorer and Einstein Telescope type improvements, which typically require entirely new facilities, \textit{will} greatly improve the low-frequency band of the detector noise leading to impressive improvements in our ability to test negative PN order GR deviations. 

\vspace{0.6cm}
\fbox{\parbox{8cm}{
{\bf Negative PN GR deviations are best constrained by gravitational waves produced by widely separated binaries, while positive PN deviations are roughly independent of the system considered.}
}}
\vspace{0.0cm}

GR deviations that enter first at negative PN order terms (relative to the leading GR term) scale with inverse powers of the gravitational wave frequency (and thus the orbital frequency). Since inspiral signals have a chirping nature, the low frequency part of a gravitational wave signal corresponds to a large separation of the binary. An example of this is the GW150914 event which, although observed with aLIGO during the very late inspiral and merger phase, would have been observed with a space-based detector when it was very widely separated. Therefore, GR deviations that enter first at negative PN order are best constrained by widely separated binaries. On the other hand, GR deviations that enter first at positive PN order are constrained equally well by all compact binary systems, provided the late inspiral and merger is in band.

\vspace{0.6cm}
\fbox{\parbox{8cm}{
{\bf Sensitivity modifications at low frequencies greatly improve our ability to constraint GR deviations that first enter at negative PN order, but modifications at high frequency do not improve positive PN constraints as much as sensitivity modifications in the bucket of the band.}
}}
\vspace{0.0cm}
 
This is because negative PN order modifications are very large at low frequencies. Improving the noise sensitivity in this regime leads to the accumulation of many more cycles at low-frequencies, and thus, to the build up of more signal-to-noise ratio precisely in the regime of the band where the modifications are largest. This can be quantified in terms of the \emph{effective cycles} accumulated at low-frequencies~\cite{Sampson:2014qqa}.
 
\vspace{0.6cm}
\fbox{\parbox{8cm}{
{\bf  Both space and ground-based detectors can place constraints that are comparable to, and sometimes better than, current constraints, though the latter can typically do somewhat better than the former.}
}}
\vspace{0.0cm}

Future gravitational wave observations will certainly lead to constraints that are in many cases more stringent than current constraints, as seen in Table~\ref{tab:summary2}. Though the results are theory-dependent, space-based instruments can often offer more stringent constraints on the properties of nature in the extreme-gravity regime, such as the mass of the graviton or the size of a large extra-dimension. This is due in part to the high signal-to-noise ratio nature of detections that are accessible to these kinds of detectors, as well as the wide range of binary masses, separation distances, and luminosity distances that produce mHz frequency gravitational waves.

\vspace{0.6cm}
\fbox{\parbox{8cm}{
{\bf  Future ground-based detectors are complementary to space-based detectors when placing constraints on modified theories of gravity.}
}}
\vspace{0.0cm}

As shown in Table.~\ref{tab:summary2}, the constraints one can place on modified gravity with ground- and space-based detectors are not significantly different. What is important is that the constraints derived with either type of instrument are, in many cases, orders of magnitude stronger than current bounds obtained by other observations and experiments. In this sense, the science case for the next generation of ground-based instruments and for space-based instruments is strong with regards to the inferences one can extract about theoretical physics from future gravitational wave data. 


The remainder of this paper describes in detail the methodology used to reach the results summarized above and is divided as follows. 
Section~\ref{sec:mod-GR} explains how different modifications to the pillars of GR imprint onto the gravitational wave observable. 
Section~\ref{sec:data-analysis} presents the data analysis tools and gravitational wave models we employ in this paper. 
Section~\ref{sec:tests-of-GR} describes the projected constraints we will be able to place on deviations from GR with future observations. 
Section~\ref{sec:inferences} maps these constraints to inferences we can extract on fundamental theoretical physics. 
Section~\ref{sec:conclusions} concludes and points to future research.  
Henceforth, we follow the conventions of~\cite{MTW}. In particular, the metric signature is $(-,+,+,+)$, Latin and Greek letters in index lists stand for parameter and spacetime indices respectively, and we use geometric units in which $G=1=c$.

\section{Modifications to the pillars of GR}
\label{sec:mod-GR}

Modified theories of gravity have pervaded the realm of gravitational physics for ages. However, we are now in a position to begin to test these competing hypotheses against actual data in the extreme gravity regime. Rather than focus our study on a particular theory of gravity, we take the alternative viewpoint of attempting to learn about and constrain deviations in the pillars upon which GR rests, agnostic to any particular theory. In this section, we classify modified gravity effects by the main (i.e. leading-order in the inspiral) deviations they impose on the pillars of GR, separating them into two groups of deviations: those that affect the generation of gravitational waves and those that affect the propagation of gravitational waves. We then discuss how such deviations imprint in the gravitational wave observable. We do not present here all possible modifications to GR pillars, and instead summarize a few important modifications following the more comprehensive analysis of~\cite{Yunes:2016jcc}.

\subsection{Modifications in the Generation of Gravitational Waves}
Modifications in the generation of gravitational waves are active only at times when the time derivatives of the multipole moments of the spacetime that generates the gravitational waves are non-zero. For a binary system, this means that generation modifications are only active during the coalescence event, whose duration depends on the total mass of the system: at most $\sim$100 minutes for stellar-mass binaries, but longer than the lifetime of space-based instruments for extreme mass-ratio events. Clearly then, generation modifications depend on the \emph{local} properties of the binary, and not on global quantities like the distance of the source to Earth.     

\subsubsection{Presence of Dipole Radiation}
Far from the source, gravitational waves can be described through a multipolar decomposition, known as a post-Minkowskian expansion, i.e.~an expansion about Minkowski spacetime in the strength of the gravitational field~\cite{Blanchet:2006zz}. In Einstein's theory and to leading-order in this expansion, gravitational waves are generated by the second derivative of the quadrupole moment of the matter source; the monopole and dipole terms do not generate gravitational waves due to conservation of mass and linear momentum, which in turn arise due to the conservation of the stress-energy tensor. In GR, then, compact binaries generate predominantly quadrupolar gravitational waves, which then carry predominantly quadrupolar energy away from the source, forcing the binary to inspiral at a given rate. 

In several modified gravity theories, however, additional scalar and vector fields can activate in regimes of extreme gravity, leading to additional sinks of energy that force binaries to inspiral faster than in GR. This typically comes about because these additional fields do not satisfy a conservation law, i.e.~their stress-energy does not satisfy an equivalent version of matter stress-energy conservation and therefore can have a monopolar structure far from the source which then forces them to carry dipolar radiation as they propagate out to spatial infinity~\cite{Stein:2013wza}. For example, in scalar-tensor theories~\cite{Fierz:1956zz,Brans:1961sx,Will:1989sk,Yunes:2011aa,Mirshekari:2013vb,Damour:1992we,Damour:1993hw}, dipole radiation is activated in the presence of neutron stars due to the excitation of a scalar field, which causes binary neutron stars to inspiral faster than predicted in GR.

Different modified theories predict that different types of binaries activate a scalar field. In scalar-tensor theories~\cite{Fierz:1956zz,Brans:1961sx,Will:1989sk,Yunes:2011aa,Mirshekari:2013vb,Damour:1992we,Damour:1993hw}, no-hair theorems~\cite{Hawking:1972qk,Sotiriou:2011dz} guarantee that black holes will not activate a monopolar scalar field, assuming a constant background field\footnote{If the background scalar field is not constant, black holes can grow hair as shown in~\cite{Horbatsch:2011ye,Healy:2011ef}}, therefore black hole binaries will not lose energy to dipole emission and will not inspiral at a dipolar rate. In quadratic gravity theories~\cite{Yagi:2015oca,Yagi:2012vf,Yunes:2011we,Yagi:2011xp,Alexander:2009tp,Jackiw:2003pm}, however, the scalar field is sourced by curvature invariants, so black holes can activate a monopole field. In such theories then, black hole binaries will lose energy and inspiral faster than predicted in GR.  Since we wish to remain theory agnostic, in this paper we will consider the activation of dipolar radiation for any compact binary systems, regardless of whether it contains neutron stars or black holes.  

How does such dipole energy loss affect the gravitational wave observable? The response function detected by instruments on Earth is the projection of the gravitational wave metric perturbation onto a response tensor. The Fourier transform of the former can be computed in the stationary-phase approximation (SPA) by integrating a function that depends on the energy loss rate. Let us then parameterize the latter for a quasi-circular compact binary inspiral via
\be
\dot{E} = \dot{E}_{\GR} + \delta \dot{E}_{\dip} v^{-2}\,,
\ee
where $\dot{E}_{\GR}$ is the energy loss rate predicted in GR, $v$ is the relative orbital velocity, and $\delta \dot{E}_{\dip}$ is a dipole correction to the GR prediction. The latter is in principle a dimensionless combination of the parameters of the system (like the masses and spins) and the coupling constants of the theory. Notice that the dipole correction to the energy loss rate is $v^{-2}$ larger than the leading GR prediction, as expected from a dipolar correction. Such a modification in the energy flux leads to the following leading PN order correction to the Fourier phase of the gravitational wave metric perturbation~\cite{Yunes:2016jcc}:
\be
\Psi_{\GW,\dip} = \Psi_{\GW,\GR} - \frac{3}{224} \eta^{2/5} \delta\dot{E}_{\dip} \left( \pi {\cal{M}}_{z} f\right)^{-7/3}\,,
\ee
where $\Psi_{\GW,\GR}$ is the Fourier phase of the gravitational wave metric perturbation in GR (see e.g.~\cite{Husa:2015iqa,Khan:2015jqa} and references therein), ${\cal{M}}_{z} = (1+z) {\cal{M}}$ is the redshifted chirp mass, ${\cal{M}} = \eta^{3/5} m$ is the source chirp mass, $\eta = m_{1} m_{2}/m^{2}$ is the symmetric mass ratio, $m = m_{1} + m_{2}$ is the total mass, and $f$ is the observed gravitational wave frequency.
	
\subsubsection{Anomalous Accelerations, Large Extra Dimensions and Time-Varying Fundamental Constants}
Modified theories that attempt to reconcile quantum mechanics with GR sometimes posit the existence of extra dimensions (in addition to the four spacetime dimensions of GR). In string theory, these extra dimensions are typically compactified and small, but in the late 1990s researchers began to consider the possibility of large extra dimensions~\cite{Randall:1999ee,Randall:1999vf,Yunes:2009bv,Sefiedgar:2010we}. For example, in the Randall-Sundrum braneworld scenario~\cite{Randall:1999ee,Randall:1999vf} (which we will consider in this paper), four-dimensional spacetime resides on a (4-dimensional) \emph{brane}, with a large extra dimension orthogonal to it leading to the \emph{bulk}. 

Until recently, it was not clear whether stable black hole solutions exist in such braneworld models. Initially, it was believed that stable solutions do not exist, and thus, that brane-localized black holes evaporate at a rate dictated in part by the mass of the black hole and the size of the large extra dimension~\cite{emparan-conj,tanaka-conj}. Recently, however, brane-localized black hole solutions have been found~\cite{Figueras:2011gd,Abdolrahimi:2012qi,Wang:2016nqi}, therefore it is possible that classical black holes do not need to evaporate in such models. Nonetheless, if black holes were to evaporate, they would force a binary containing at least one black hole to acquire an anomalous acceleration (since the mass would become time-dependent), leading to a clear signature in the gravitational wave observable~\cite{Yagi:2011yu}.   

An anomalous acceleration due to an evaporating mass is equivalent to a time-varying gravitational constant $G$~\cite{Yunes:2009bv}. This is because the binding energy of a binary, which controls the acceleration, depends on the product of the total mass of the binary and the gravitational constant, where the latter acts as a conversion factor between mass and energy~\cite{Yunes:2009bv}. Scalar tensor theories~\cite{Fierz:1956zz,Brans:1961sx} can be thought of as promoting the gravitational constant to a function of a spacetime-dependent scalar field in the presence of matter, while $F(R)$ theories~\cite{Frolov:2011ys} and bimetric theories~\cite{1973PhRvD...8.3293L} can introduce a time-dependence in $G$ in pure vacuum spacetimes. One can then imagine a situation in which the mass is constant but $G$ is a slowly-varying function of time, which would then lead to an anomalous acceleration and modification to the gravitational wave observable that is similar to that induced by an evaporating braneworld black hole. 

The promotion of $G$ to a spacetime quantity violates local position invariance, one of the key components of a fundamental pillar of GR, the Strong Equivalence Principle (SEP).  The SEP states that the trajectory of any body (weakly- or self-gravitating) is not only independent of its internal structure and composition, but also of the velocity of the (freely-falling) frame in which one measures this trajectory and of when and where in the Universe the object is located; this last component of the SEP is called local position invariance. Therefore, a constraint on the anomalous acceleration caused by a time-variation of $G$ or of $M$ is a test of the SEP. 
 
How does this anomalous acceleration affect the gravitational wave observable? In the case of an evaporating black hole binary, the leading PN order correction to the Fourier phase of the gravitational wave metric perturbation is~\cite{Yagi:2011yu,Yunes:2016jcc}:
\begin{align}
\Psi_{\GW,\ED} &= \Psi_{\GW,\GR} + \frac{25}{851968} \; \dot{m}  
\nonumber  \\
&\times
\frac{3 - 26 \eta + 34 \eta^{2}}{\eta^{2/5} (1-2\eta)}  \left( \pi {\cal{M}}_{z} f\right)^{-13/3}
\end{align}
where, as before, $\Psi_{\GW,\GR}$ is the Fourier phase of the gravitational wave metric perturbation in GR, while $\dot{m} \equiv dm/dt$ is the source's evaporation rate which depends on the size of the extra dimension. Of course, this is only valid for binaries that contain black holes, since these contain singularities through which gravitons can, in principle, leak into the bulk of a higher dimensional model. In the case of a time-dependent gravitational constant, the Fourier phase is~\cite{Yunes:2009bv,Yunes:2016jcc}:
\begin{align}
\Psi_{\GW,\dot{G}} &= \Psi_{\GW,\GR} - \frac{25}{65526} \frac{\dot{G}_{z} {\cal{M}}_{z}}{G}  \left( \pi {\cal{M}}_{z} f\right)^{-13/3}
\end{align}
where $\dot{G}_{z} = \dot{G}/(1 + z)$ is the observed rate of change of Newton's constant. In both cases, the correction to GR depends on the same power of frequency, and one can map $\Psi_{\GW,\ED}$ to $\Psi_{\GW,\dot{G}}$ through a redefinition of $\dot{G}$ in terms of $\dot{m}$.

\subsubsection{Local Lorentz Symmetry Violation}

The SEP requires that local Lorentz symmetry be preserved by the gravitational interaction, i.e.~that the outcome of all gravitational experiments be independent of the velocity of the (freely-falling) laboratory frame in which the measurements are performed. Violations of this symmetry are present when one introduces preferred frames aligned with dynamical vector fields, as is the case in Einstein-\AE{}ther theory~\cite{Jacobson:2000xp,Jacobson:2008aj} or when one choses a preferred foliation of spacetime through a dynamical scalar field, as in a (healthy) version of Ho\v{r}ava-Lifshitz gravity~\cite{Horava:2008ih,Horava:2009uw,Vacaru:2010rd,Blas:2011zd} called khronometric gravity~\cite{Blas:2009qj,Blas:2010hb}. 

The magnitude of the violation of Lorentz symmetry is controlled by the magnitude of the coupling constants of the theory~\cite{Yagi:2013qpa,Yagi:2013ava,Hansen:2014ewa}. In Einstein-\AE{}ther theory, the degree of Lorentz violation depends on four coupling constants $(c_{1},c_{2},c_{3},c_{4})$, two of which are stringently constrained by Solar System experiments, leaving only two, namely $c_{\pm} = c_{1} \pm c_{3}$, that are weakly constrained. In khronometric gravity, the degree of violation is controlled by three coupling constants $(\alpha_{\KG},\beta_{\KG},\lambda_{\KG})$, the first of which is stringently constrained by Solar System experiments, leaving only the latter two weakly constrained. The best constraints on these two degrees of freedom in both Einstein-\AE{}ther theory and khronometric gravity comes from binary pulsar observations~\cite{Yagi:2013qpa,Yagi:2013ava}.

How do gravitational Lorentz-violations affect the gravitational wave observable? In Einstein-\AE{}ther theory, the leading PN order correction to the Fourier phase of the gravitational wave metric perturbation is~\cite{Hansen:2014ewa}:
\begin{align}
\Psi_{\GW,\EA} &= \Psi_{\GW,\GR} +\frac{3}{128}\left( \pi {\cal{M}}_{z} f\right)^{-5/3} \left[1  
- \left( 1-\frac{c_{14}}{2} \right) 
\right. \nonumber \\
&\times \left.
\left(\frac{1}{w_2}+\frac{2c_{14}c_+^2}{(c_++c_--c_-c_+)^2 w_1} +\frac{3c_{14}}{2w_0 (2-c_{14})} \right) 
\right. \nonumber \\
&+ \left. {\cal{S}} {\cal{A}}_{2} + {\cal{S}}^{2} {\cal{A}}_{3}
\right]
\end{align}
while in khronometric gravity one finds~\cite{Hansen:2014ewa}:
\begin{align}
\Psi_{\GW,\KG} &= \Psi_{\GW,\GR}+\frac{3}{128}(\pi \scM_{z} f)^{-5/3}\bigg[1-\left(1-\beta_{\KG}\right)
\bigg. \nonumber \\
&\times \left.
\left(\frac{1}{w_{2}^{\KG}} + \frac{3\ \beta_{\KG}}{2 w_{0}^{\KG}(1-\beta_{\KG})}\right)
\right. \nonumber \\
&+ \left. {\cal{S}} {\cal{A}}_{2} + {\cal{S}}^{2} {\cal{A}}_{3}
\right]
\end{align}
where $w_{n}$ is the propagation speed of the spin-$n$ mode, ${\cal{S}} = (s_{1} m_{2} + s_{2} m_{1})/m$, with $s_{1,2}$ the sensitivities of the compact objects\footnote{The sensitivities of the compact objects are essentially a measure of how the mass changes as the scalar field varies. See e.g.~\cite{Will:1994fb,Eardley1,Will:1977wq}.}, and ${\cal{A}}_{2,3}$ given in~\cite{Yagi:2013qpa,Yagi:2013ava}. Notice that both corrections enter at \emph{Newtonian} order, meaning they are both proportional to $(\pi {\cal{M}} f)^{-5/3}$ which is the leading PN order dependence on frequency of $\Psi_{\GW,\GR}$. In principle, there is also a correction that enters at $-1$PN order, but this is proportional to the difference in the sensitivities of the compact object, which have not yet been calculated for black holes in either theory. In Sec.~\ref{sec:tests-of-GR}, we will neglect this -1PN order correction, thus obtaining conservative projected constraints, i.e.~constraints that could in principle become more stringent if the -1PN term were included once the sensitivities are calculated. 

\subsection{Modifications in the Propagation of Gravitational Waves}
Modifications in the propagation of gravitational waves are active only while the wave travels from the source to Earth. Since all sources are at cosmological distances, billions of light years away, there is ample time for these modifications to compile during their travel to Earth. This means, in particular, that propagation modifications are proportional to a positive power of the source's distance to Earth, a \emph{global} quantity, and thus, they can typically be many orders of magnitude larger than generation modifications (assuming both types are present). Typically, propagation modifications are enhanced relative to generation modifications by a factor of the ratio of the times during which each of these effects is active, i.e.~$\sim t_{\rm prop}/t_{\rm gen} \sim D_{L}/{\cal{M}}$, where $D_{L}$ is the luminosity distance and ${\cal{M}}$ is the total mass~\cite{Yunes:2016jcc}.  

In GR, the gauge boson that carries the gravitational interaction, the graviton, is massless which need not be the case in modified theories. In massive gravity theories~\cite{Will:1997bb,Rubakov:2008nh,Hinterbichler:2011tt,deRham:2014zqa}, the gravitational interaction is mediated by a massive gauge boson that must travel slower than the speed of light. Using insight from special relativity, one then expects that the (phase) velocity of a massive graviton $v_{g}$ should satisfy
\begin{equation}
\frac{v_{g}^{2}}{c^{2}}=1-\frac{m_{g}^{2}c^{4}}{E^{2}},
\end{equation}
where $c$ is the speed of light, $m_{g}$ is the rest mass of the graviton, and $E$ is its rest energy ~\cite{Will:2014kxa}. If $m_{g}>0$, then $\frac{v_{g}}{c}<1$ and the velocity of the gravitational wave will be slower than the speed of light. In the limit $m_{g}\to 0$, one of course recovers the predictions of Einstein's theory.

Such a modification in the graviton's dispersion relation will then propagate into a correction to the gravitational wave observable. To all PN order, the modified Fourier phase is~\cite{Will:1997bb,Rubakov:2008nh,Hinterbichler:2011tt,deRham:2014zqa,Yunes:2016jcc}:
\begin{align}
\Psi_{\GW,\MG} &= \Psi_{\GW,\GR}  + \pi^{2} \frac{D_{0}}{(1+z)} \frac{{\cal{M}}_{z}}{\lambda_{\MG}^{2}} \left(\pi {\cal{M}}_{z} f\right)^{-1}
\end{align} 
where $D_{0}$ is a measure of the distance to the source, ${\cal{M}}_{z}$ is the redshifted chirp mass, $z$ is the cosmological redshift and $\lambda_{\MG}$ is the Compton wavelength of the massive graviton~\cite{Will:1997bb}. Notice that, as expected, the massive gravity modification is proportional to the distance traveled by the wave.

In addition to a propagation modification, massive gravity theories may also modify the generation of gravitational waves, although this depends much more on the particular massive gravity model considered. As demonstrated in~\cite{Yunes:2016jcc}, however, when both a modification to the generation and to the propagation of gravitational waves are present, the latter will dominate the former, i.e.~the ratio of a propagation modification to a generation modification scales as $D_{L}/{\cal{M}}$.

\section{Data Analysis, Models, \\ Detectors and Sources}\label{sec:data-analysis}

Gravitational waves interact very weakly with matter, which is why they are typically buried in detector noise. If one knows the shape of the gravitational wave one expects to detect, then the optimal search strategy is \emph{matched filtering}, when the noise is stationary and Gaussian. In this strategy, one maximizes the cross-correlation of the detector's output and a waveform response model or template (weighted by the detector's spectral noise density) with respect to the template's parameters to find the best fit. In this section, we will first describe the gravitational wave models we employ, and then continue with a discussion of our data analysis strategy and a description of future detectors through their spectral noise density. 
\subsection{Gravitational Wave Models to Test GR}\label{subsec:waveform-model}

A matched filtering strategy to test GR with gravitational waves requires the use of waveform templates that represent the gravitational wave response function in modified gravity, the first step of which is the construction of accurate templates \emph{within} GR. In this paper, we use the inspiral, merger, and ringdown (IMR) templates of~\cite{Khan:2015jqa, Santamaria:2010yb,Husa:2015iqa}, sometimes referred to as the (non-precessing) \emph{PhenomD} model; we refer the reader to those references for a detailed description of how such waveform models are constructed. These waveforms are appropriate to describe the gravitational waves emitted in the inspiral, merger, and ringdown of binary black holes. Indeed, PhenomD was constructed by fitting the plunge and merger phase to numerical relativity simulations of approximately equal-mass black hole coalescences. In spite of this, we will continue to use PhenomD when modeling the inspiral of binaries with at least one neutron star component, as well as when modeling extreme mass-ratio inspirals. When considering the former, we will be forced to stop all of calculations before the merger occurs, i.e.~at the gravitational wave frequency corresponding to first contact         since the merger is typically drastically different. When considering the latter, one should in principle include higher PN order corrections to the inspiral phase. In this paper, however, we will use the PhenomD model as a {\emph{kludge waveform}} when considering extreme- and intermediate-mass ratio systems, as was done in the past e.g.~in~\cite{Barack:2003fp}. This kludge treatment should be enough to obtain a correct qualitative understanding of the constraints we can place with such systems; when more accurate waveforms are computed for these systems, the conclusions of this paper can be refined.

For the waveform model in modified gravity, we use the parameterized post-Einsteinian (ppE) framework~\cite{Yunes:2009ke}, a nested super-model built on top of a given GR model. This framework modifies any given GR waveform model through the introduction of two classes of \emph{theory parameters}: \emph{exponent parameters} (that specify the particular type of GR deviation) and \emph{amplitude parameters} (that control the magnitude of the GR deformation). Gravitational wave detectors are much more sensitive to the gravitational wave phase than to their amplitude, and thus, neglecting amplitude modifications for simplicity, the \emph{simplest} ppE  model is of the form~\cite{Yunes:2009ke}
\begin{equation}
\tilh_{\ppE}(f) =\tilh_{\GR}(f) e^{i\beta u^{b}}
\label{eq:ppE}
\end{equation}
where $u=(\pi \scM f)^{1/3}$ is a reduced frequency, $\beta$ is a ppE amplitude parameter, and $b$ is a ppE exponent parameter. 

Lacking numerical relativity intuition of how gravitational waves are modified in the merger phase of a binary coalescence, we will only consider modifications to the inspiral part of the waveform. An analysis similar to what is done in this paper was done in \cite{Berti:2016lat} but for ringdown modes. The ppE inspiral-merger-ringdown waveform becomes
\begin{equation}\label{hppE} 
\tilh_{\ppE}=\begin{cases}
		\tilh_{\GR}^{\rm ins} e^{i\beta u^{b}} & f < f_{\IM},\\
		\tilh_{\GR}^{\MR} & f >  f_{\IM},\\
		\end{cases}
\end{equation}
where $\tilh_{\GR}^{\rm ins}$ and $\tilh_{\GR}^{\MR}$ are the GR Fourier waveforms in the inspiral and in the merger-ringdown respectively, and $f_{\IM}$ is the frequency of transition from the inspiral phase to the merger-ringdown phase, with the fitting parameters of the model chosen to ensure continuity of the waveform across $f_{\IM}$. As previously stated, we will model the GR part of the waveform with the IMRPhenomD model. 

The ppE model is particularly useful because it not only allows us to constrain generic deviations from GR, it also allows us to map these parameter constraints to bounds on physical processes that modify the pillars of Einstein's theory.  In particular, different values of $b$ correspond to different physical mechanisms that introduce such modifications, as should be clear by comparing Eq.~\eqref{hppE} to the Fourier phase corrections presented in Sec.~\ref{sec:mod-GR} (see also~\cite{Yunes:2016jcc}). Table~\ref{PN Table} presents a simple reference to connect the deviations presented in Sec.~\ref{sec:mod-GR} to ppE parameters.
\begin{table*}[htb]
\begin{tabularx}{\textwidth}{Y|Y|M|Y|M}
\hline\hline
Modified GR Pillar & Example Theories & $b$ &$\beta$& PN  \\ \hline
Dipole Radiation	& 	Scalar Tensor, EdGB 	&	-7	&  $\beta_{\mbox{\tiny ST}}$~\cite{Horbatsch:2011ye,Yunes:2011we}, $\beta_{\mbox{\tiny EdGB}}$~\cite{Yagi:2011xp}&	-1 \\
Local Position Invariance (SEP)	& 	Extra Dimension~\cite{Randall:1999ee,Randall:1999vf,Yunes:2009bv,Sefiedgar:2010we}, $G(t)$	&  	-13	& $\beta_{\mbox{\tiny ED}}$~\cite{Yagi:2011yu}, $\beta_{\dot{G}}$~\cite{Yunes:2009bv}	&-4  \\
Local Lorentz Violation (SEP)	& 	Einstein-\AE{}ther, Khronometric  	&	-5	&  $\beta_{\EA}$~\cite{Jacobson:2000xp,Jacobson:2008aj}, $\beta_{\KG}$~\cite{Blas:2009qj,Blas:2010hb}&	0 \\
\hline
Massive Graviton  	& 	 Massive Gravity	&	 -3	&	 $\beta_{\MG}$~\cite{Will:1997bb,Rubakov:2008nh,Hinterbichler:2011tt,deRham:2014zqa}&+1 \\
Modified Dispersion Relation  	& 	 Quantum Gravity	&	 $3\alpha-3$	&	 $\beta_{\MDR}$~\cite{Yunes:2016jcc}&$1+\frac{3}{2}\alpha$ \\
\hline\hline
\end{tabularx}
\caption{\label{PN Table} Mapping between violations to the pillars of GR (first column), a few example modified theories in which such violations occur (second column), and the ppE parameters that recover such modifications (third and fourth column), together with the leading PN order at which they enter (last column). The table is divided into those modifications that arise during the generation of gravitational waves (top) and those that appear during the propagation of gravitational waves (bottom)~\cite{Yunes:2016jcc}.}
 \end{table*}

Given a gravitational wave observation that is consistent with GR, one can then determine how large $\beta$ can be while remaining consistent with statistical noise. To do so, one must maximize the cross-correlation between the signal, which we will assume to be given exactly by the waveform evaluated in GR (i.e.~the ppE model with $\beta = 0$), and the ppE model with respects to all parameters of the latter. The parameters of the ppE model are $\theta^{a}=(\Ln \, \scA, \phi_{c}, t_{c}, \Ln \, \scM, \Ln \, \eta, \chi_{s}, \chi_{a}, \beta )$, where $\scA$ is an amplitude factor that scales with $\scM_{z}^{5/6}/D_{L}$, $\scM_{z}$ is the redshifted chirp mass, $D_{L}$ is the luminosity distance, $\phi_{c}$ is the phase of coalescence, $t_{c}$ is the time of coalescence, and $\chi_{s,a} \equiv (\chi_{1} \pm \chi_{2})/2$ are symmetric and anti-symmetric combinations of the (dimensionless) spin parameters $\chi_{1,2} \equiv \vec{S}_{1,2}/m_{1,2}^{2}$, where $\vec{S}_{1,2}$ is the spin angular momentum of the compact object. We do not include the polarization angle or the sky location angles in the parameter vector, as we are using sky-averaged waveforms.

\subsection{The Basics of a Fisher Analysis}

The most accurate way to determine how stringently modified gravity deviations can be constrained with future observations is through a Bayesian analysis. In such a study one calculates the full posterior probability distribution of the search parameters given the observations. The width of such posteriors then provides a measure of how much statistical wiggle room there is for parameters that represent GR deviations to vary from zero.  

For sufficiently high signal-to-noise ratio signals~\cite{Vallisneri:2007ev,Vallisneri:2011ts}, an approximation to this Bayesian calculation, a Fisher analysis, provides an upper bound on constraints when testing GR by the Cramer-Rao bound~\cite{Cramer,Rao}. In a Fisher analysis, one assumes the likelihood probability function has a single Gaussian peak, and approximates the behavior of the signal about that peak through a Taylor expansion. What results is a measure of the variance and the covariance of parameters in the template model through integrals that depend only on the templates and the spectral noise density of the detector. In what follows, we summarize the main details of this calculation, following the notation of~\cite{Will:1994fb,Heavens:2009nx,Berti:2004bd}.
\begin{table*}[tb]
\begin{tabularx}{1\textwidth}{Y|Y|Y|Y|Y|Y|Y}
   \hline\hline
Name& $m_{1} [M_{\odot}]$ & $m_{2} [M_{\odot}]$ & $(\chi_{1},\chi_{2})$ & $D_{L}$ & z & $\rho_{\mbox{\tiny LISA}}$\\\hline
 GW150914 &  35.1   	& 	  29.5  	&	(0.31, 0.39)	& 	400 Mpc 	& $\sim$ 0.09 & 6.6	\\
EMRI &  $10^{5}$ 	& 	10 	&	(0.8, 0.4)	&	1 Gpc	& $\sim$ 0.2	&102.2	\\
IMRI &$10^{5}$ 	& 	 $10^{3}$	&	(0.7, 0.9)	& 5 Gpc & $\sim$ 0.8	&297.5	\\
IMBH & 5$ \times 10^{3}$ 	& 	4$\times10^{3}$  	&	(0.7, 0.9)	&	 16 Gpc		& $\sim$ 2& 	102.7	\\
SMBH &  5$\times10^{6}$ 	& 	4$\times10^{6}$ 	&	(0.7, 0.9)	&	 48 Gpc	&$\sim$ 5 &	486.7\\
   \hline\hline
\end{tabularx}
\caption{\label{table:LISA} Intrinsic (source) properties of one representative system considered for space-based detectors, chosen to be representative of various classes that could emit gravitational waves in the frequency-range of these detectors. The luminosity distances were chosen to conform with astrophysical expectations and are fixed across detectors. The signal-to-noise ratios vary depending on the instrument used to evaluate this statistic, and here we present only those provided by the LISA configuration.}
\end{table*}
Given a waveform model $h(t;\theta^{a})$ with parameter $\theta^{a}$, the root-mean-squared ($1\sigma$) error on any single parameter in a Fisher analysis is given by
\be
\label{eq:1sigma}
\Delta \theta^{a}=\sqrt{\Sigma^{aa}},
\ee
where no sum is here implied and where the variance-covariance matrix $\Sigma^{ab}$ is found by inverting the Fisher matrix $\Gamma_{ab}$, i.e.~$\Sigma^{ab}=(\Gamma_{ab})^{-1}$. The Fisher matrix is given by
\begin{equation}
\label{eq:Fisher}
\Gamma_{ab}\equiv \bigg( \frac{\sd h}{\sd \theta^{a}}\bigg|\frac{\sd h}{\sd \theta^{b}}\bigg),
\end{equation}
where $\sd h/ \sd \theta^{a}$ is the partial derivative of the waveform model with respect to the parameter $\theta^{a}$, and where we have defined the inner product between two waveform models via
\begin{equation}
(h_{1}|h_{2})\equiv 2 \int_{f_{\low}}^{f_{\hi}}\frac{\tilh_{1}\tilh_{2}^{*}+\tilh_{1}^{*}\tilh_{2}}{S_{n}(f)}\ df\,.
\end{equation}
In the inner product, the overhead tilde stands for the Fourier transform, the superscript star stands for complex conjugation, and $S_{n}(f)$ is the detector's spectral noise density. With this definition of the inner product, the signal-to-noise ratio is simply 
\begin{equation}
\rho \equiv (h|h)^{1/2}.
\end{equation}

The lower and upper limits of integration in the inner product can be effectively taken to be the frequencies at which the noise of the detector becomes very large. For space-based detectors, we choose $f_{\low}$ to be
\be
f_{\low}^{\spc} = \max\left(f_{\ratiolo}, f_{\yrs}\right)\,,
\ee
where $f_{\ratiolo}$ is the (low) frequency at which the ratio of the amplitude of the gravitational wave signal to the noise spectrum is 1:10, while $f_{\yrs}$ is the frequency that corresponds to three years prior to reaching the inner most stable circular orbit frequency, $f_{\isco}$, for a test particle in a Schwarzschild spacetime. Similarly, we choose $f_{\hi}$ to be
\be
\label{eq:fhi-sp}
f_{\hi}^{\spc} = f_{\ratiohi}\,,
\ee
where $f_{\ratiohi}$ is the (high) frequency at which the gravitational wave amplitude is once again overcome by the noise by a ratio of 1:10. For ground-based detectors, we choose $f_{\low}$ to be 
\be
f_{\low}^{\grnd} = \max\left(f_{\locut}, f_{\ratiolo} \right)\\,
\ee
where $f_{\locut} = 5 \; {\rm{Hz}}$ when considering aLIGO at design sensitivity, A+, Voyager, and CE, while $f_{\locut} = 1 \; {\rm{Hz}}$ when considering ET. Similarly, we choose $f_{\hi}$ to be 
\be
\label{eq:fhi-gr}
f_{\hi}^{\grnd} = \min\left(f_{\ratiohi},f_{\hicut}\right)\,,
\ee
where $f_{\rm \hicut} = f_{\rm cont}$ is the gravitational wave frequency at contact when considering binaries with at least one neutron star, and $f_{\rm \hicut} = f_{\ratiohi}$ when considering binary black holes. We stop the integration at the contact frequency when considering neutron stars because the waveform model we use does not incorporate features present in neutron star mergers (see Sec.~\ref{subsec:waveform-model}) 

With this machinery in hand, in Sec.~\ref{sec:tests-of-GR} we will present estimates of the projected accuracy to which GR deviations can be constrained through a Fisher analysis on ppE PhenomD models, with the amplitude multiplied by the square root of the number of detectors (one for both CE and ET-D, and two for aLIGO, A~+, Voyager, and the space-based detectors). The noise curves used in the analysis are sky-averaged, as discussed in Sec.~\ref{subsec:future-dets}. All calculations are carried out in \textit{Mathematica}, making sure that the numerical accuracy of all calculations is high enough to be accurate to at least percent level. In particular, the Fisher matrix is inverted through a Cholesky decomposition to ameliorate the approximate singular nature of the former. The Fisher analysis is carried out to constrain $\beta$ for ppE exponent parameters $b\in \{-13, -12, ... , -1\}$, equivalent to modified gravity corrections of $n\in\{-4, -3.5, ... , +2\}$ PN order. One could in principle consider a wider range of values for $b$, but this prior range suffices to capture all of the physical processes presented in Sec.~\ref{sec:mod-GR}.  

\subsection{Future Detectors and their Sources}\label{subsec:future-dets}
This subsection describes a few possible detector configurations that have been considered by the community and the sources we expect to detect with them. We will not provide a detailed technical description of each of the future instruments, but instead we characterize them via their spectral noise density. The latter will effectively determine how well parameters can be measured with a Fisher analysis. For more details about the detectors, we refer the reader to~\cite{Klein:2015hvg,LIGO-Tech-doc,Audley:2017drz}
\subsubsection{Space-based Detectors and their Sources}\label{sec:fut-detc}
\begin{figure}[t]
\begin{center}
\includegraphics[width=\columnwidth,clip=true]{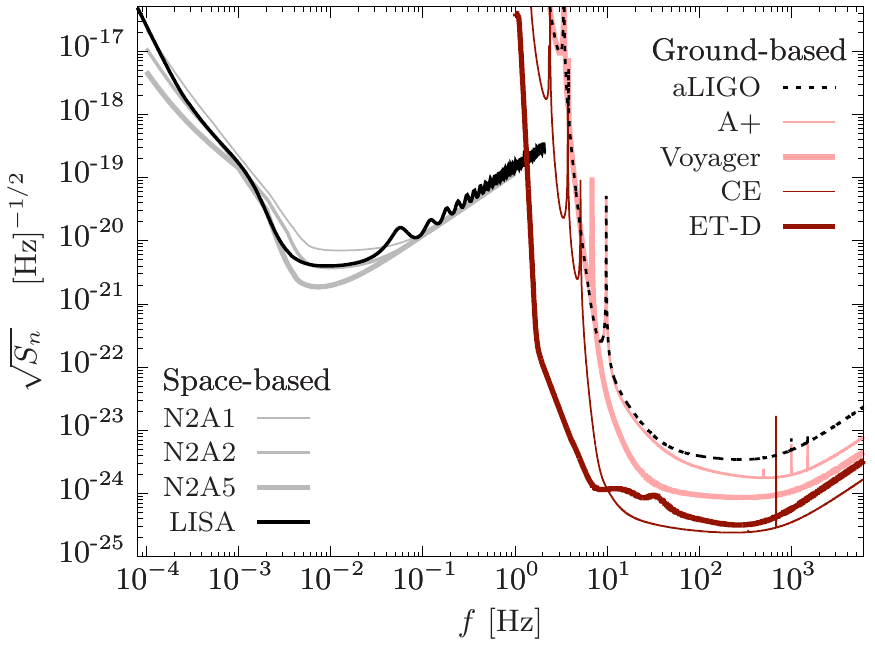}
\caption{\label{fig:Sncurves}(Color Online) Noise spectral density for all instruments used in our analysis as a function of frequency. (e)LISA operates at much lower frequencies than aLIGO. At such low frequencies, (e)LISA is capable of detecting both gravitational waves from very massive systems inaccessible to aLIGO, as well as the very early inspiral of some systems whose merger-ringdown phases occur in the aLIGO frequency band, such as GW150914-like systems.}
\end{center}
\end{figure}

A proposal for the final design of LISA has recently been submitted for review by ESA~\cite{Audley:2017drz}.
This instrument will have six-links with 2.5 Gm arms and low acceleration noise demonstrated possible with LISA Pathfinder~\cite{Armano:2016bkm}.
We also consider three other previously suggested eLISA configurations with different sky-averaged, six-link sensitivity curves presented in~\cite{Klein:2015hvg} that differ only in the length of the arms (1, 2, and 5 Gm corresponding to the labels A1, A2, and A5). We only consider configurations with low acceleration noise; these correspond to the N2 configurations of~\cite{Klein:2015hvg}. Figure~\ref{fig:Sncurves} presents the spectral noise densities for each LISA configuration we consider as a function of frequency. Each successive eLISA configuration improves the sensitivity of the instrument in the low- and middle-frequency regions, with N2A5 being the most sensitive configuration (the ``classic LISA'' design). All throughout we will assume a three-year mission duration.

\begin{table*}[tb]
\begin{tabularx}{1\textwidth}{Y|Y|Y|Y|Y|Y|Y|Y|Y|Y}
   \hline\hline
Name& $m_{1} [M_{\odot}]$ & $m_{2} [M_{\odot}]$ & $(\chi_{1},\chi_{2})$ & $D_{L}$ &  $\rho_{\mbox{\tiny aLIGO}}$&$\rho_{\mbox{\tiny A+}}$&$\rho_{\mbox{\tiny Voyager}}$&$\rho_{\mbox{\tiny ET-D}}$&$\rho_{\mbox{\tiny CE}}$\\\hline
NSNS&  2 	& 	1.4 	& 	(0.01, 0.02) 		& 	 100 Mpc	& 23.2 & 33.6 &109.5 &238.4& 382.7	\\
$\ell$BHNS & 5 	& 	1.4 	&	(0.2, 0.02)	& 	150 Mpc	&21.7 & 31.3 & 102.8 & 225.1 & 361.2	\\
$\ell$BHBH & 8 	& 	5 	&	 (0.2, 0.3) 	&	250 Mpc	&27.7 & 39.8 & 131.4 & 289.0 & 463.7	\\
BHBH & 25 	& 	20  	&	 (0.3, 0.4) 		&	 800 Mpc  & 21.7 & 28.4 & 108.2 & 253.4 & 409.1 \\
GW150914&  35.1 	& 	29.5 	&	 (0.31, 0.39)  		&	 400 Mpc & 54.6 & 71.5 & 271.8 & 641.7 & 1031.1	\\
   \hline\hline
\end{tabularx}
\caption{\label{table:aLIGO} Same as Table~\ref{table:LISA} but for one representative system appropriate to \emph{ground-based} detectors. The distances were fixed such that the signal-to-noise ratio is always $\sim25$ for an aLIGO detection at design sensitivity, except for the distance to (and other properties of) the GW150914 system which was fixed to be approximately that of the actual GW150914 event.}
\end{table*}

\allowdisplaybreaks[4]
For each LISA configuration, we consider gravitational waves emitted by the following classes of systems: 
\begin{itemize}
\item \emph{GW150914-like systems (GW150914):}  A low mass black hole binary with moderate spins, low redshift, and a mass ratio of $q \equiv m_{2}/m_{1} \sim0.8$.
\item \emph{Extreme mass-ratio inspirals (EMRIs):} A low-mass black hole (with a mass between $10-100M_{\odot}$) inspiralling into a supermassive black hole (with a mass between $10^{5}-10^{7}M_{\odot}$) with moderate to high spins and a relatively low redshift.
\item \emph{Intermediate mass-ratio inspirals (IMRIs):} An intermediate-mass black hole (with a mass between $10^{3}-10^{4}M_{\odot}$) inspiralling into a supermassive black hole (with a mass between $10^{5}-10^{7}M_{\odot}$) with moderate to high spins and redshifts of order unity. 
\item \emph{Intermediate mass black hole binaries (IMBH):} Two intermediate-mass black holes with masses between $10^{3}-10^{5}M_{\odot}$, moderate to high spins, and at moderate redshift.
\item \emph{Supermassive black hole binaries (SMBH):} Two supermassive black holes with masses between $10^{6}-10^{7}M_{\odot}$, moderate to high spins, moderate to high redshifts, and a high mass ratio.  
\end{itemize}
All systems are assumed to be on a quasi-circular inspiral trajectory. The luminosity distance is fixed individually for each system, with all distances corresponding to a redshift smaller than 10. Notice that since the signal-to-noise ratio scales with $\scA$, which is proportional to the chirp mass, these signal-to-noise ratios are typically much larger than the signal-to-noise ratios of events that will be detected by ground-based instruments. For each of these classes, we pick three representative systems to explore projected constraints on deviations from GR. One representative of each class is listed in detail in Table~\ref{table:LISA}. In this paper, we do not consider constraints obtained by stacking multiple events, as this will be studied separately in the future~\cite{stacking-future}.

Since detections with space-based instruments will have very high signal-to-noise ratios, it becomes important to consider not only statistical error in our parameter estimations, but also systematic error. Systematic error becomes dominant to statistical error for sufficiently loud signals which will have the effect of saturating our projected bounds. However, as detectors improve, waveform modeling will also become more accurate, and thus, systematic error will also decrease. We expect the results of this study to be roughly insensitive to the inclusion of additional complexity in the inspiral waveforms as more accurate models become available. 

\subsubsection{Ground-based Detectors}
\begin{figure*}[htb]
\begin{center}
\includegraphics[width=1\textwidth,clip=true]{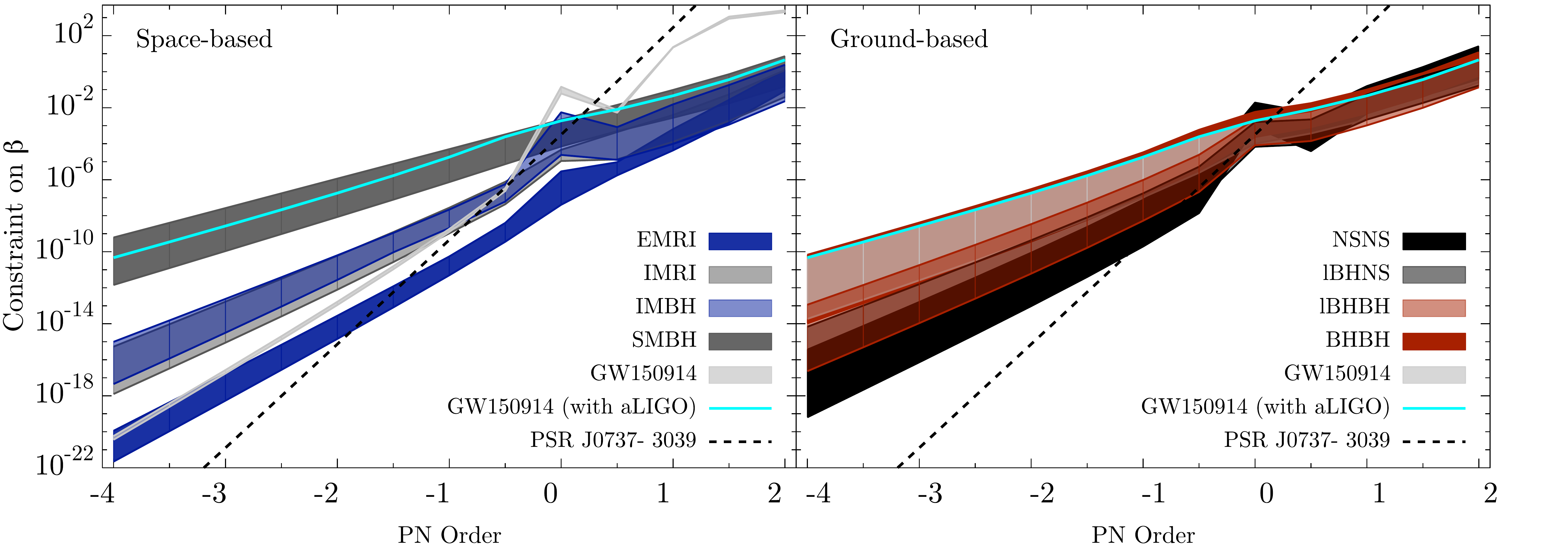}
\caption{\label{shaded}(Color Online) Projected constraints on modified gravity effects as a function of ppE PN order at which they first enter for a variety of space-based (left) and ground-based (right) detectors and a variety of systems (anything above the regions is projected to be ruled out). The shaded regions are bounded by the highest and lowest constraints that can be placed at a given PN order for all instruments studied. For comparison, we also include the constraints that have already been placed by aLIGO with the GW150914 detection~\cite{PhysRevLett.116.061102,Abbott:2016nmj} (thin cyan line), as well as constraints that can be placed with binary pulsars~\cite{Yunes:2010qb} (dashed black line). Observe that the magnitudes of the projected constraints with space-based and ground-based instruments are comparable, with the former being better by roughly $2$--$4$ orders of magnitude at negative PN order.}
\end{center}
\end{figure*}

We consider a set of spectral noise density curves that correspond to modeled estimates of the noise in various instrument realizations: aLIGO (at design sensitivity), A+, Voyager, CE, and ET-D~\cite{LIGO-Tech-doc,Punturo:2010zz,Matt-priv}. The second and third are currently projected as aLIGO upgrades, while the last two are new, third-generation ground-based detectors. Some key properties of these future instruments are the following:
\begin{itemize}
\item \emph{A+.} Projected date of operation of $\sim 2020$, improves the level of quantum noise and coating thermal noise, increasing the observational range of aLIGO by 140\%.
\item \emph{Voyager.} Projected date of operation of $\sim 2027$, reduces the aLIGO noise by using silicon in place of glass in mirrors and suspensions, as well as operating at a lowered temperature of 120K (rather than aLIGO's 295K), increasing the observational range of A+ by a factor of roughly two. 
\item \emph{CE:} Projected date of operation of $\sim 2035$, will be a new facility that is much larger than aLIGO and possibly underground, increasing the observational range of aLIGO by a factor of 10 to 100. 
\item \emph{ET-D:} With a projected date of operation of $\sim 2030$-$2035$, will be a new facility built underground to decrease the low-frequency noise, thus increasing the observational range of aLIGO by roughly the same amount as CE. 
\end{itemize}
Figure~\ref{fig:Sncurves} presents the spectral noise density curves for each of these instruments.

For each ground-based detector, we consider gravitational waves emitted by the following classes of systems:  
\begin{itemize}
\item \emph{Neutron star binaries (NSNS):} A neutron star binary system with negligible spins at very low redshift.
\item \emph{Low-mass black hole-neutron star binaries ($\ell$BHNS):} A neutron star inspiraling into a stellar-mass black hole with small spins at very low redshift.  
\item \emph{Low-mass black hole binaries ($\ell$BHBH):} A stellar-mass ($5-10M_{\odot}$) black hole binary system with small to moderate spins at small redshift. 
\item \emph{Black hole binaries (BHBH):} A black hole binary system with masses in the tens of solar masses, small to moderate spins, and at small redshift.
\item \emph{GW150914-like binary systems (GW150914):} As previously defined in Sec.~\ref{sec:fut-detc}.
\end{itemize}
As in the space-based case, all systems are assumed to be on a quasi-circular inspiral trajectory. The luminosity distances for the ground-based systems are chosen such that each system has a signal-to-noise ratio of $\sim$25 when detected with aLIGO at design sensitivity, except for the GW150914 case which uses the actual detection parameters. As expected, at a fixed luminosity distance, the signal-to-noise ratios increase as the instrument sensitivity improves. As in the space-based case, for each of these classes we pick three representative systems to explore projected constraints on deviations from GR (the properties of one of these is listed in Table~\ref{table:aLIGO}). As in the case of sources for space-based detectors, we will not consider constraints obtained by stacking multiple events, leaving this to future work~\cite{stacking-future}.

\begin{figure*}[htb]
\begin{center}
\includegraphics[width=\columnwidth,clip=true]{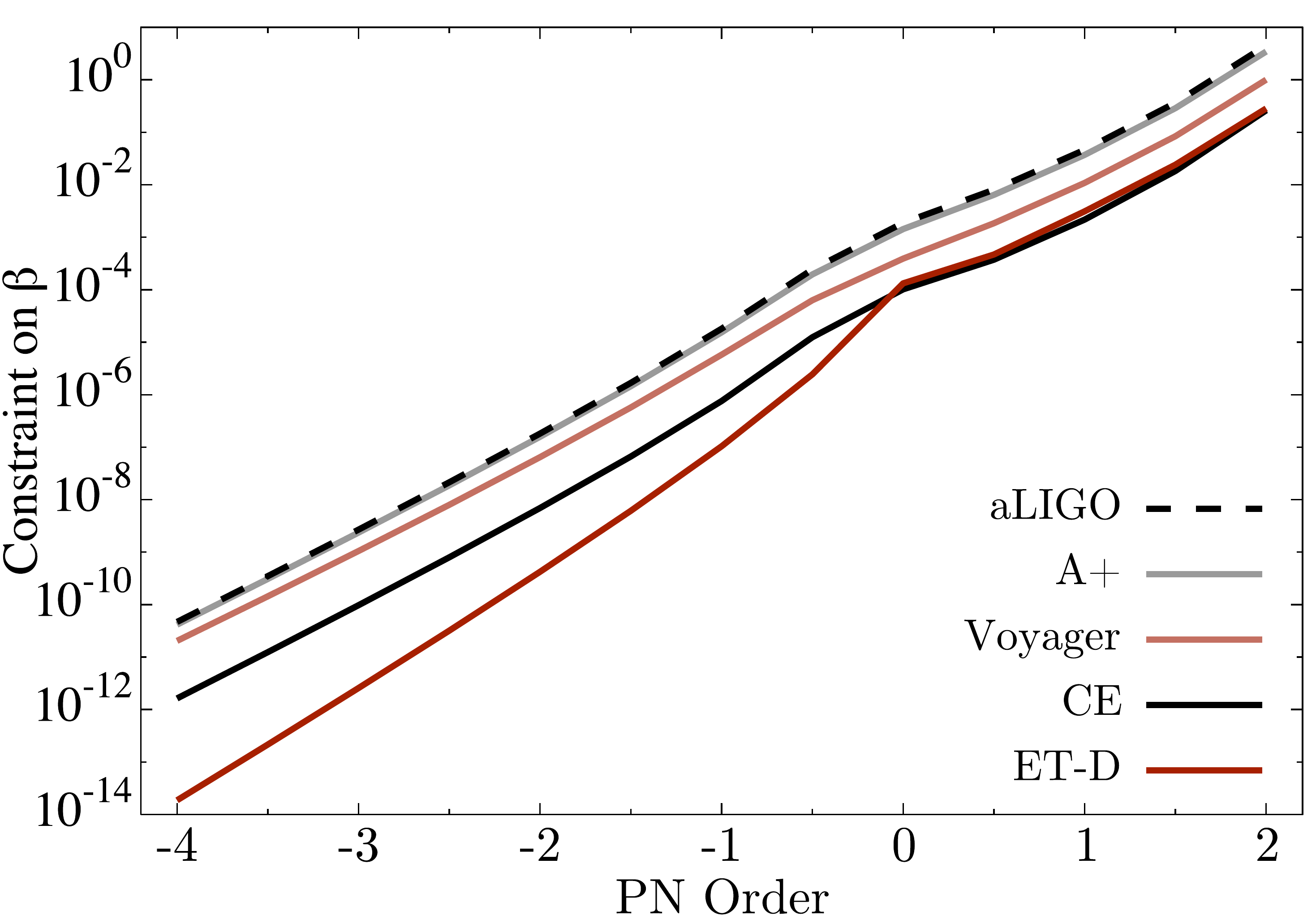} \quad
\includegraphics[width=\columnwidth,clip=true]{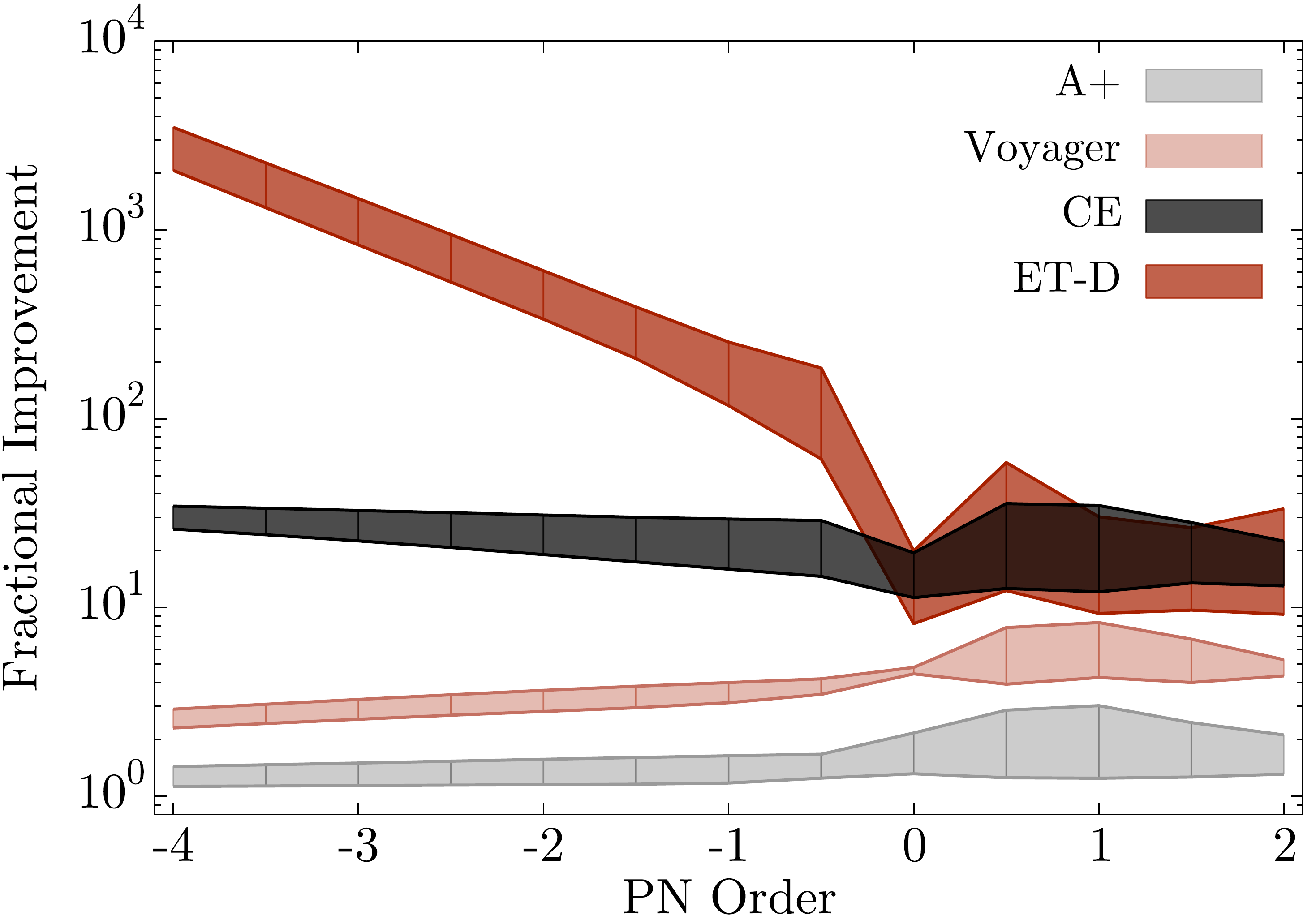}
\caption{\label{fig:mods}(Color Online)~
Left: Projected constraints on $\beta$, given a GW150914-like detection with different ground-based detectors. Observe that ET-D provides better constraints than CE for all negative PN orders due to its low-frequency improvement, but they become comparable for PN orders $\geq 0$. 
Right: Fractional improvement to the projected constraints on GR modifications relative to those achievable with aLIGO at design sensitivity, as a function of PN order. The boundaries of the shaded regions are determined by the minimum and maximum constraints that can be placed with an instrument when considering all systems under study. Observe that while A+ improves the constraints only marginally relative to aLIGO, Voyager and CE can provide constraints that are roughly 5 and 20 times better respectively. Observe also that ET-D leads to drastically better constraints relative to aLIGO at negative PN orders, but this improvement becomes comparable to that obtained with CE at positive PN orders, as also shown on the left panel.}
\end{center}
\end{figure*}

\section{Prospects for Future Tests of GR}\label{sec:tests-of-GR}

This section discusses how projected constraints on deviations from GR are improved with future detectors. We begin by presenting these constraints and conclude with an explanation of the improvements using phenomenological noise curves. 
\subsection{Future Ground-based and Space-based Tests}\label{sec:results}
The left panel of Fig.~\ref{shaded} shows the projected constraints that space-based detectors can place on modified gravity as a function of the ppE PN order at which the modification first enters for a variety of systems (the region above the curves would be ruled out given such observations). The constraints are presented as shaded regions, which represent variation due to instrument choice and representative system choice. In all cases, the N2A5 configuration can do best at testing GR, as expected from Fig.~\ref{fig:Sncurves}. For negative PN order modifications, the SMBH class is worst at placing constraints, while the EMRI and GW150914 classes are best. For positive PN order modifications, all classes do approximately equally well, except for the GW150914 class, which does the worst.  

Let us explain this behavior. Negative PN order modifications to GR are proportional to negative powers of the orbital velocity (relative to the leading PN order GR term). Therefore, negative PN corrections are naturally larger for systems that are more widely separated since their orbital velocity is smaller by the Virial theorem (a version of which typically holds in modified gravity). The GW150914-like systems that space-based detectors could observe are by far the most widely separated and, therefore, lead to the best projected constraints at negative PN order. EMRI systems could also be emitting gravitational waves during the entire lifetime of the space-based missions at relatively small velocities, although not as small as those of GW150914-like systems. Nonetheless, the constraints one could place with EMRIs at negative PN orders are comparable to those we can place with GW150914-like systems because (i) we have chosen signal-to-noise ratios an order of magnitude larger for the former and (ii) the gravitational waves emitted by these EMRI systems start in the bucket of the LISA noise curve (at about $5 \times 10^{-3}$ Hz). 

At positive PN order, the projected constraints on modifications to GR one could place seem to be roughly independent of the system considered. Positive PN corrections are largest in the last stages of the inspiral of binary systems, and in particular, during their merger. But the maximum velocity reached by all these systems is roughly the same before merger. This, combined with the different signal-to-noise ratios of the different systems explains why they roughly achieve the same constraints. Of course, this is not the case for the GW150914-like systems, since these do not merge in the LISA band.   

The projected constraints presented here are clearly better than \emph{current} constraints we can place with the aLIGO GW150914 and GW151226 observations (even if this had been detected at design sensitivity). This is particularly true at negative PN order, where constraints can be ten orders of magnitude more stringent with future LISA observations. Such a result was already anticipated in~\cite{Barausse:2016eii}, where constraints on dipole radiation were investigated. Indeed, our results are consistent with those in~\cite{Barausse:2016eii}, extending them to other PN orders. 

Comparing constraints on modified gravity that can be achieved with space-based detectors in the future to those we can place with aLIGO today is unfair. By the time LISA flies, it is very probable that new and improved versions of aLIGO will be operational. For this reason, the right panel of Fig.~\ref{shaded} shows projected constraints as a function of PN order using aLIGO (at design sensitivity) and all future ground-based detectors for a variety of systems. As expected, systems that sample the smallest orbital velocities (NS binaries in the ground-based case) will lead to the most stringent constraints on negative PN order effects, while all systems do approximately the same at positive PN order. Comparing the best constraints on the left and the right panels, we also see that LISA can place constraints that are roughly 3 orders of magnitude better than the best future ground-based constraints at negative PN order, while the constraints are roughly the same at positive PN order.   

Another feature we notice by comparing the left and right panels of Fig.~\ref{shaded} is that the shaded regions are much wider in the ground-based case than in the space-based case. This is because although there are several possible LISA configurations, they differ little relative to future improvements to ground-based instruments. For this reason, it is instructive to look at the fractional improvement in the projected constraints with future upgrades to ground-based instruments relative to the constraints we can obtain with aLIGO at design sensitivity as a function of PN order at which they enter. This is shown in Fig.~\ref{fig:mods} where the fractional improvement is defined via
\be
{\rm{Frac.}} \; {\rm{Improv}} = \frac{{\rm{Future}} \; {\rm{Det.}}}{{\rm{aLIGO}} \; {\rm{Design}}}\,, 
\ee
with the shaded regions representing variability in the constraints due to the different systems that could be detected. Observe that the A+ and Voyager detectors improve the constraints we can place on modified gravity at all PN orders by a factor of about $2$--$10$ respectively. On the other hand, the ET and CE detectors can lead to drastic improvements in our ability to test GR, with enhancements of up to two orders of magnitude at negative PN order and one order of magnitude at positive PN order. Observe that CE is better than ET at positive PN orders, while the reverse is true at negative PN orders, which is due to the improved low-frequency sensitivity of ET, as shown in Fig.~\ref{fig:Sncurves}.

The constraints obtained with binary pulsars (such as J0737-3039~\cite{Yunes:2010qb}) appear to do noticeably better than gravitational wave constraints for all negative PN orders. However, as one considers more advanced ground-based detectors and a larger class of systems, gravitational waves can do better than binary pulsars for modifications that enter at PN orders $\gtrsim -1$. Similarly, in the space-based case, gravitational waves can do better than binary pulsars for modifications that enter at PN orders $\gtrsim -2$. 

When comparing binary pulsar constraints to gravitational wave constraints, caution is advised. Binary pulsar constraints obviously require spacetimes with matter, since the observations depend on radio pulses emitted by neutron stars. Many modified theories of gravity, however, modify GR predominantly in vacuum spacetimes, almost exactly reducing to GR near stars, regardless of their compactness. For example, in Einstein dilaton Gauss-Bonnet gravity~\cite{Yunes:2011we,Yagi:2011xp} and in dynamical Chern-Simons gravity~\cite{Yunes:2009hc,Yagi:2013mbt}, black holes acquire non-trivial scalar hair, but the scalar hair of neutron stars is greatly suppressed~\cite{Yagi:2015oca}. Therefore, binary pulsar observations cannot constrain these classes of theories at all, and thus, it is important to make observations involving at least one black hole (be it with an as-of-yet undetected radio pulsar observation from a black hole-neutron star binary, or with gravitational wave observations of coalescing black hole or black hole-neutron star binaries).

\begin{figure*}[htb]
\begin{center}
\includegraphics[width=.5\textwidth,clip=true]{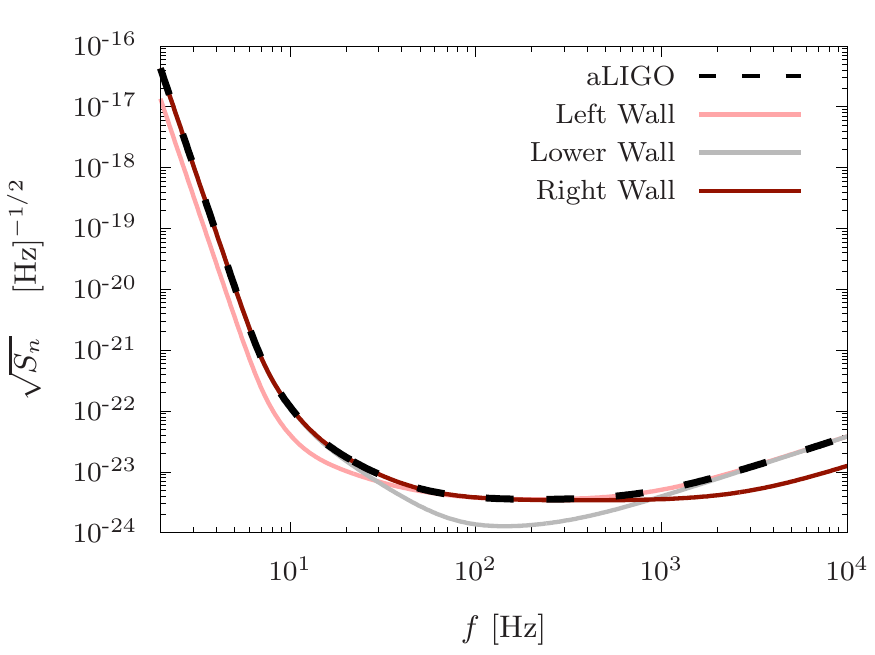}
\includegraphics[width=.455\textwidth,clip=true]{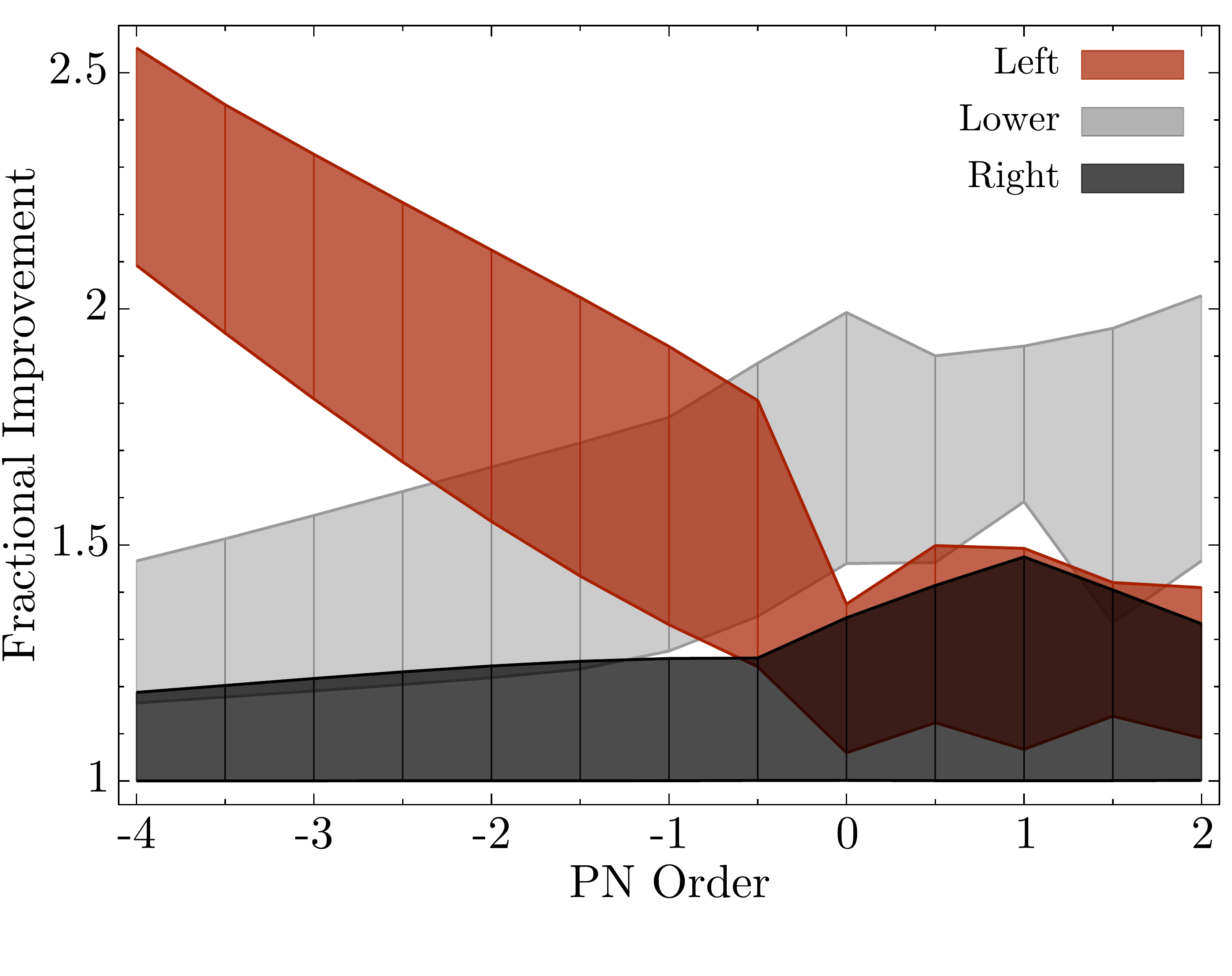}
\caption{\label{fig:phenom}(Color Online) Left: Phenomenological noise spectral density curves as a function of frequency. Right: Relative fractional improvement of the constraints on $\beta$ provided by the phenomenologically modified sensitivity curve over those provided by a fit to design aLIGO. Observe that the noise curve improved at low-frequencies leads to the largest improvements at negative PN order, while the opposite is true for the noise curves improved in the mid- and high-frequency ranges.}
\end{center}
\end{figure*}

\subsection{Exploration of Detector Design}

In the previous subsection, we provided an intuitive explanation of why certain sensitivity improvements impact certain PN order modifications more than others. Here, we verify these explanations by investigating constraints with \emph{phenomenological} noise curves that isolate improvements to the sensitivity in the low-, mid-, and high-frequency regimes.  

Let us begin by presenting the phenomenological noise curves we consider. We constructed an analytic fit to the design-aLIGO spectral noise density (zero-detuned, high-power) noise curve~\cite{Matt-priv} that resembles in functional form the one presented in~\cite{Mishra:2010tp}. We then divide the frequency range over which aLIGO operates into three domains: the low-frequency regime ($f<50 \mbox{Hz}$), the mid-frequency regime ($50\mbox{Hz}<f<900\mbox{Hz}$), and the high-frequency regime ($f>900\mbox{Hz}$). We improve the sensitivity by roughly a factor of ten without modifying the other domains, leading to a ``left wall'' improvement (low-frequency regime), a ``lower wall'' improvement (mid-frequency regime), and a ``right wall'' improvement (high-frequency regime), as seen on the left panel of  Fig.~\ref{fig:phenom}. 

The right panel of Fig.~\ref{fig:phenom} shows the relative fractional improvement in the constraints we can place on modified gravity with these phenomenological noise curves as a function of PN order. Observe that the left-wall improvements dominate at negative PN orders, while the lower- and right-wall improvements dominate at positive PN order. The reason for this is that as one improves the sensitivity at low-frequencies, the detector is now able to capture gravitational waves when the binary system was more widely separated and the orbital velocity was smaller, precisely where negative PN order corrections are larger (since they enter as negative powers of the velocity). Similarly, as one improves the sensitivity in the mid- and the high-frequency regimes, the detector is more sensitive to the late inspiral and merger, where the binary reaches the highest velocities and where positive PN order corrections are larger. 

One may wonder why the constraints obtained with the right-wall improvement are worse than those obtained with the lower-wall improvement. The former improves the noise curve in the high-frequency regime by definition, and thus, one may think this noise curve would be better at constraining high PN order deviations from GR. This, however, is not true, as shown in the right panel of Fig.~\ref{fig:phenom}. For any chirping signal, there are typically many more cycles at lower frequencies than at higher frequencies, and thus, more cycles accumulate in the regime improved by the lower-wall noise curve. The distinguishability of non-GR signals (as measured for example by the Bayes factor) is not simply a function of the integrated dephasing induced by the non-GR corrections, but rather is a function of the integrated \emph{noise-weighted} dephasing, which~\cite{Sampson:2014qqa} calls the \emph{effective cycles}. The lower-wall improvement leads to more effective cycles of phase for high PN order corrections than the right-wall improvement, explaining the feature we see in the right panel of Fig.~\ref{fig:phenom}.  

The improvement in the constraints is roughly the square root of the improvement in the sensitivity curve. Mathematically, this is easy to see; the constraint scales as the square-root of the variance-covariance matrix in Eq.~\eqref{eq:1sigma}, and the latter scales inversely with the spectral noise.  The improvement in the constraints is not quite a factor of 3 throughout because we have only increased the sensitivity curve in one part of the frequency domain. Physically, one can also understand the constraint improvement as a result of an increase in signal-to-noise ratio, which also scales with the square root of the spectral noise, i.e.~by increasing the sensitivity by an order of magnitude, we have roughly increased the signal-to-noise ratio by a factor of three, since we have kept the luminosity distance fixed.

The behavior observed in Fig.~\ref{fig:phenom} is reminiscent of that found in Sec.~\ref{sec:results} within the ground-based instruments. For example, A+ has the same sensitivity as design aLIGO at low-frequencies (see Fig.~\ref{fig:Sncurves}), which is why A+ cannot improve constraints on modified gravity at negative PN orders relative to what we can do with aLIGO at design sensitivity. ET-D is more sensitive than CE at low-frequencies, and thus, it allows for the most stringent constraints on negative PN order modifications to GR relative to those we can obtain with design aLIGO. Voyager's increased sensitivity in the mid and high frequencies has a fractional improvement that is similar to that obtained with the lower- and right-wall models.
\section{Theoretical Physics Implications}\label{sec:inferences}
Experimental relativity consists of more than just carrying out null-hypothesis tests and constraining generic deviations from GR. A crucial next step is to use such constraints to make inferences on modified gravity mechanisms that, since GR has been confirmed, cannot be active in the extreme gravity regime. In this section, we will map the constraints on the ppE parameters derived in the previous section to constraints on the magnitude of certain corrections to the pillars that GR rests on, as described in Sec.~\ref{sec:mod-GR}. We will enhance the study of the previous section by considering more than a single characteristic source per class, and instead consider 3 sources per class, which will allow us to show a range of possible inferences. We will then explore how these inferences change as a function of the instrument used. 
\subsection{Presence of Dipole Radiation}
\begin{figure}[tb]
\begin{center}
\includegraphics[width=\columnwidth,clip=true]{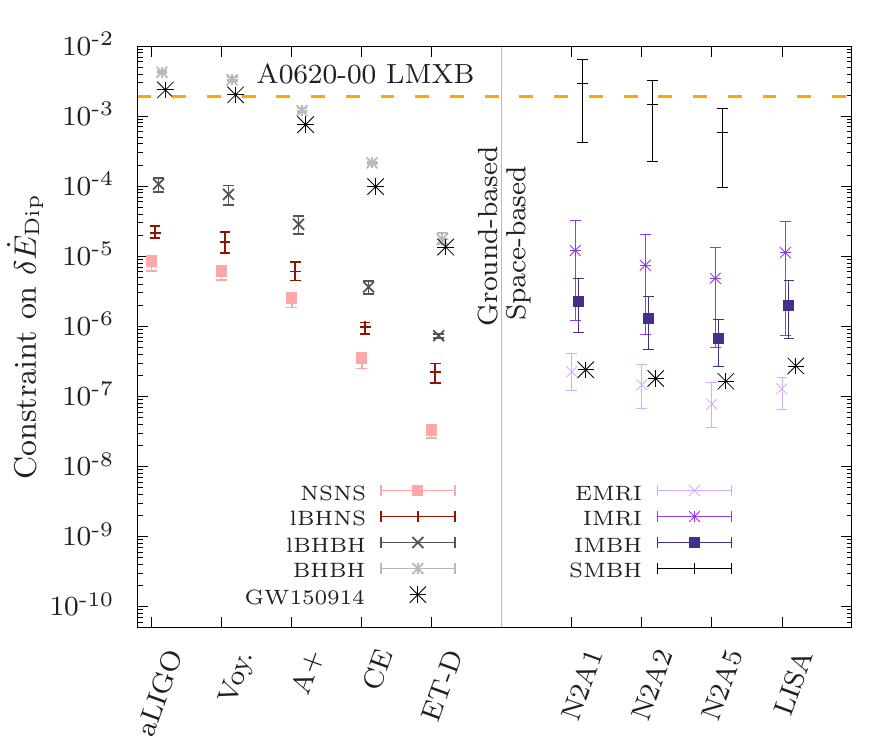}
\caption{\label{fig:dipole}(Color Online) Projected constraints on the strength of dipolar emission $\delta \dot{E}_{\dip}$ as a function of instrument. Current constraints on dipole radiation from the low mass X-ray binary pulsar are given by the horizontal dashed yellow line. The vertical lines represent the variability of the constraint within the class of systems considered. Observe the the projected constraints are 4--5 orders of magnitude stronger than the current bound (which is given by the low-mass X-ray binary~\cite{Yagi:2012gp}), and that those obtained with third-generation ground-based detectors are comparable to those obtained with space-based detectors.}
\end{center}
\end{figure}

If the orbit of compact binaries decays faster than predicted in GR due to the emission of dipole radiation, the gravitational wave phase will acquire a leading-order correction that enters at -1PN order, as described in Sec.~\ref{sec:mod-GR}. The modified waveform can then be modeled as described in Sec.~\ref{subsec:waveform-model} with the ppE mapping:
\be
\beta = -\frac{3}{224}\delta \dot{E}_{\dip} \eta^{2/5}\,,
\quad 
b = -7\,.
\ee
Therefore, for a given constraint on $\beta$, we obtain stronger constraints on $\delta \dot{E}_{\dip}$ if the signal accumulates significant SNR at lower frequencies, i.e.~when the gravitational wave producing binary is widely separated. Indeed, as seen in Fig.~\ref{fig:dipole}, the LISA configurations give the best constraints using GW150914-type and EMRI systems, while ground based detectors do best with neutron star binary systems. In both cases, the constraints are roughly 4--5 orders of magnitude stronger than the current bound obtained from observation of low-mass X-ray binaries~\cite{Yagi:2012gp}.

We also see that dipole radiation can be constrained comparably well with future ground-based and space-based instruments, although we see that the former can do better than the latter in the best case by roughly one order of magnitude. This seems to contradict the constraints on $\beta$ shown in Fig.~\ref{shaded}, which at -1PN order are roughly the same with EMRIs and neutron star binaries, the best space- and ground-based systems at this PN order respectively. The reason space-based detectors do worse is that in converting the constraint on $\beta$ to a constraint on $\delta\dot{E}_{\dip}$, one must divide by $\eta^{2/5}$ which induces a suppression in the EMRI case, but barely affects the neutron star case.  

\subsection{Anomalous Accelerations, Large Extra Dimensions and Time-Varying Fundamental Constants}
The existence and size of a single large extra-dimension introduces a leading-order modification to the gravitational wave phase that enters at -4PN order, as described in Sec.~\ref{sec:mod-GR}. In this case, however, constraints are only possible when at least one of the binary components is a black hole, as otherwise there is no leakage into the extra dimension. In this study, we only consider black hole binaries as a generalization. As before, the gravitational wave can then be modeled as in Sec.~\ref{subsec:waveform-model} with
\be
\beta = \frac{dm}{dt}\frac{25}{851968}\bigg(\frac{3-26\eta+34\eta^{2}}{\eta^{2/5}(1-2\eta)}\bigg)\,,
\quad 
b = -13\,
\ee
where $\frac{dm}{dt}\equiv\dot{m}=\dot{m_{1}}+\dot{m_{2}}$ and 

\begin{equation}
\dot{m_{a}}=-2.8\times 10^{-7}\bigg(\frac{M_{\odot}}{M_{a}}\bigg)^{2}\bigg(\frac{\ell}{10 \mu m}\bigg)^{2} M_{\odot}\ \mbox{yr}^{-1}
\end{equation}
where, in this case, $M_{a}$ is the mass of one of the black holes and $\ell$ is the size of the large extra dimension that we are interested in constraining. 

\begin{figure}[htb]
\begin{center}
\includegraphics[width=\columnwidth,clip=true]{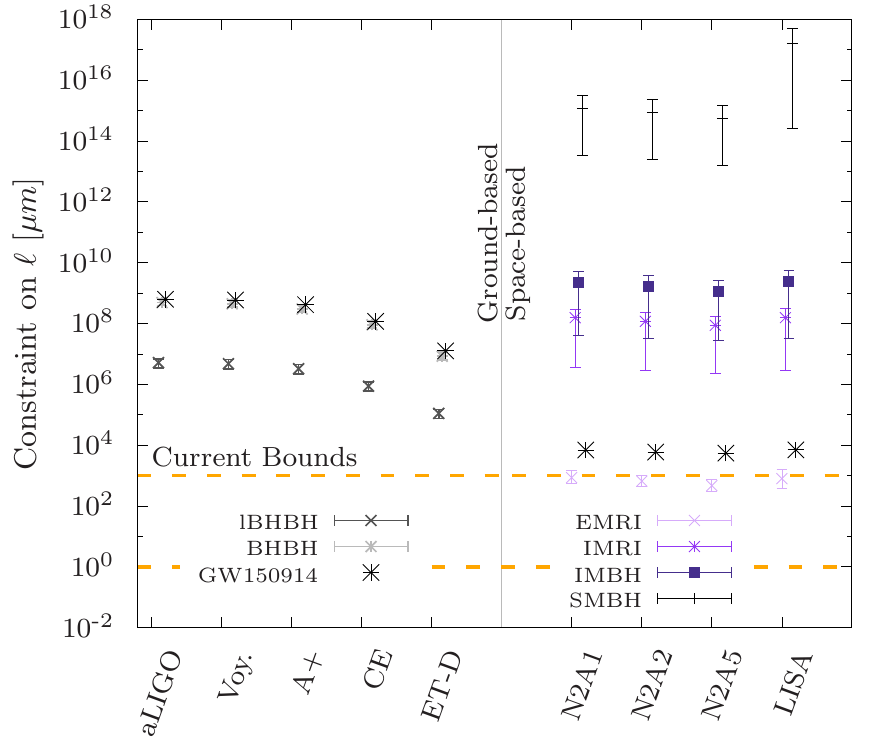}
\caption{\label{fig:ed}~(Color Online) Projected constraints on the size of a large extra dimension as a function of instrument.  Current constraints on $\ell$ are between 10 and $10^{3} \; \mu$m (see Table~\ref{tab:summary2}) as shown with horizontal dashed lines. Observe that EMRIs can place constraints on the size of a large extra dimensions that are $\sim10^{12}$ orders of magnitude more stringent than those placed with SMBHs, and that these constraints are roughly competitive with current constraints.}
\end{center}
\end{figure}

 Because the modification enters at negative PN order as in the case of dipole radiation, we expect the best constraints on $\beta$ to come from gravitational waves emitted by widely separated systems. However, the mapping above shows that any constraint on $\dot{m}$ will be enhanced by a factor of $\eta^{2/5}$, thus suggesting that the systems with most extreme masses will lead to the best constraints. Indeed, as seen in Fig.~\ref{fig:ed}, the best constraints are obtained with EMRI systems detected with space-based detectors. These constraints are approximately comparable to current constraints~\cite{Johannsen:2008tm,Johannsen:2008aa,Adelberger:2006dh,Psaltis:2006de,Gnedin:2009yt}, but 7 orders of magnitude better than the best constraints achievable with ground-based detectors. 

\begin{figure}[htb]
\begin{center}
\includegraphics[width=\columnwidth,clip=true]{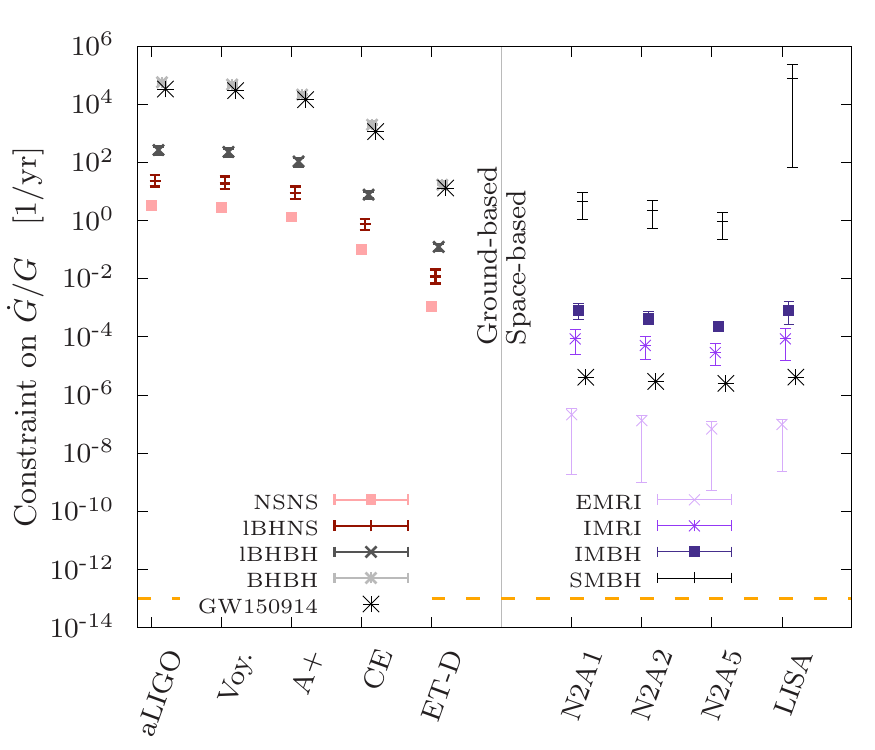}
\caption{\label{fig:gvar}~(Color Online) Projected constraints on the magnitude of the time variability of Newton's constant G as a function of instrument. Future generation detectors will be able to place constraints up to 12 orders of magnitude more stringent than design aLIGO, with space-based detec tors beating ground-based by as much as 7 orders of magnitude. Current constraints are about $10^{-13}$/yr~\cite{Bambi:2005fi,Copi:2003xd,Manchester:2015mda}, shown with a dashed horizontal line near the bottom of the figure.}
\end{center}
\end{figure}

The time-variability of Newton's constant $G$ can also be constrained by studying a -4PN order deviation from GR. As discussed in Sec.~\ref{sec:mod-GR}, the ppE mapping is in this case
\be
\beta = \frac{25}{65526}\frac{\dot{G_{z}}}{G}\scM_{z}\,,
\quad 
b = -13\,,
\ee
which then suggests that systems with large chirp mass (due to the $\beta$--$\dot{G}$ mapping) and those that are widely separated (due to the negative PN correction) will place the most stringent constraints. This is indeed reflected in Fig.~\ref{fig:gvar}, where we see the best constraints come from EMRI systems, which are 4--6 orders of magnitude better than the best constraints we can place with third-generation ground-based detectors. Since current constraints on $\dot{G}$ are $10^{-13}/yr$~\cite{Bambi:2005fi,Copi:2003xd,Manchester:2015mda}, none of these will be directly competitive.

\begin{figure}[htb]
\begin{center}
\includegraphics[width=\columnwidth,clip=true]{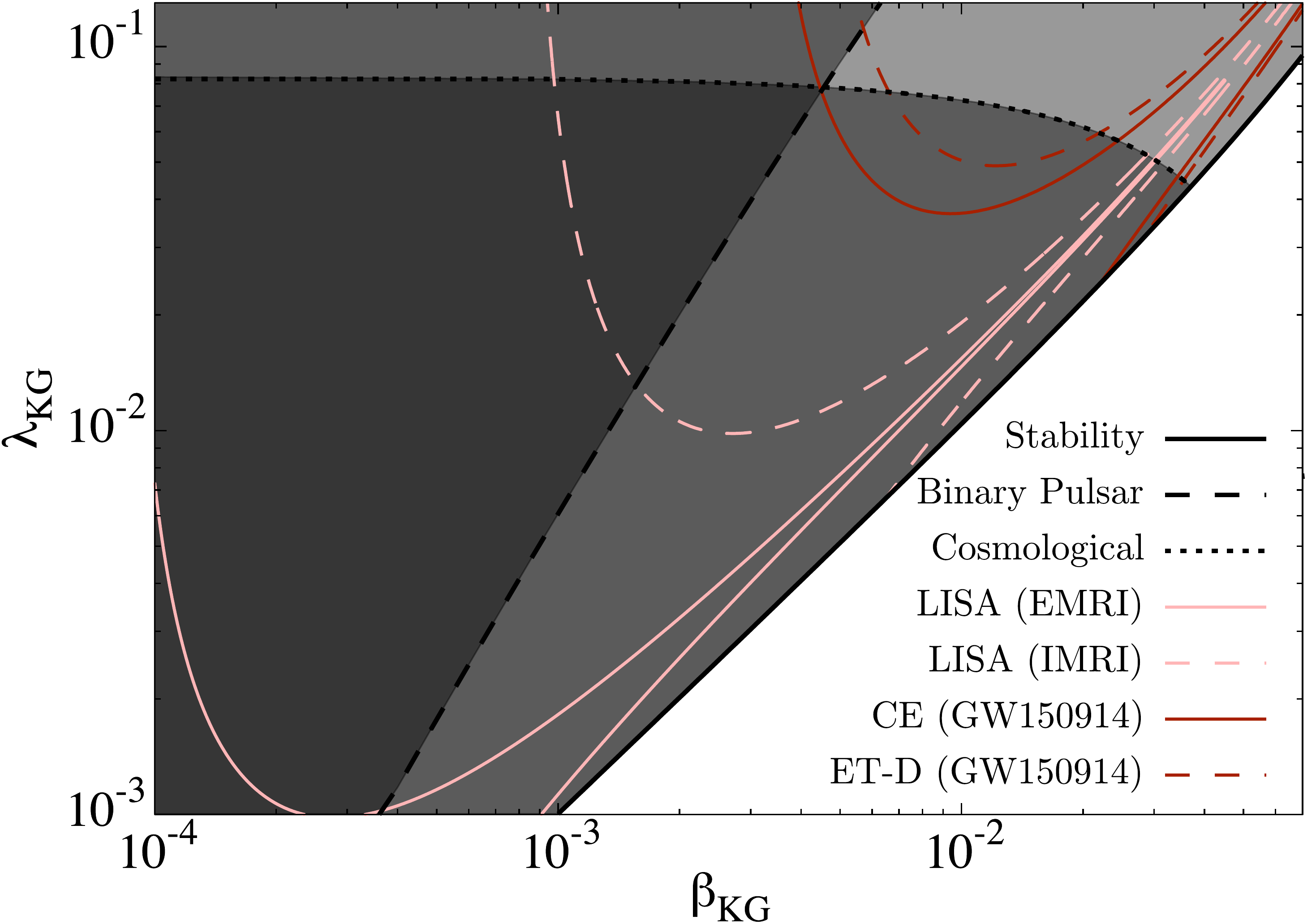}
\caption{\label{fig:KG}(Color Online) Projected constraint regions on the coupling parameters of khronometric theory. The region below the solid black line, the region to the right of the dashed black line, and the region above the dotted black line contain values of $(\beta_{KG},\lambda_{KG})$ that violate certain stability constraints~\cite{Elliott:2005va,Blas:2010hb,Barausse:2011pu}, binary pulsar constraints~\cite{Yagi:2013qpa,Yagi:2013ava}, and cosmological constraints~\cite{Jacobson:2008aj,Zuntz:2008zz,Carroll:2004ai,Audren:2013dwa} respectively. The regions above and to the right of the different color lines correspond to values of $(\beta_{KG},\lambda_{KG})$ that would be ruled out with future gravitational wave observations using different ground-based instruments. The EMRI and IMRI lines correspond to future projected constraints with LISA, while the CE and ET-D lines correspond to future projected constraints with GW150914-like observations.
}
\end{center}
\end{figure}

\begin{figure*}[htb]
\begin{center}
\includegraphics[width=\columnwidth,clip=true]{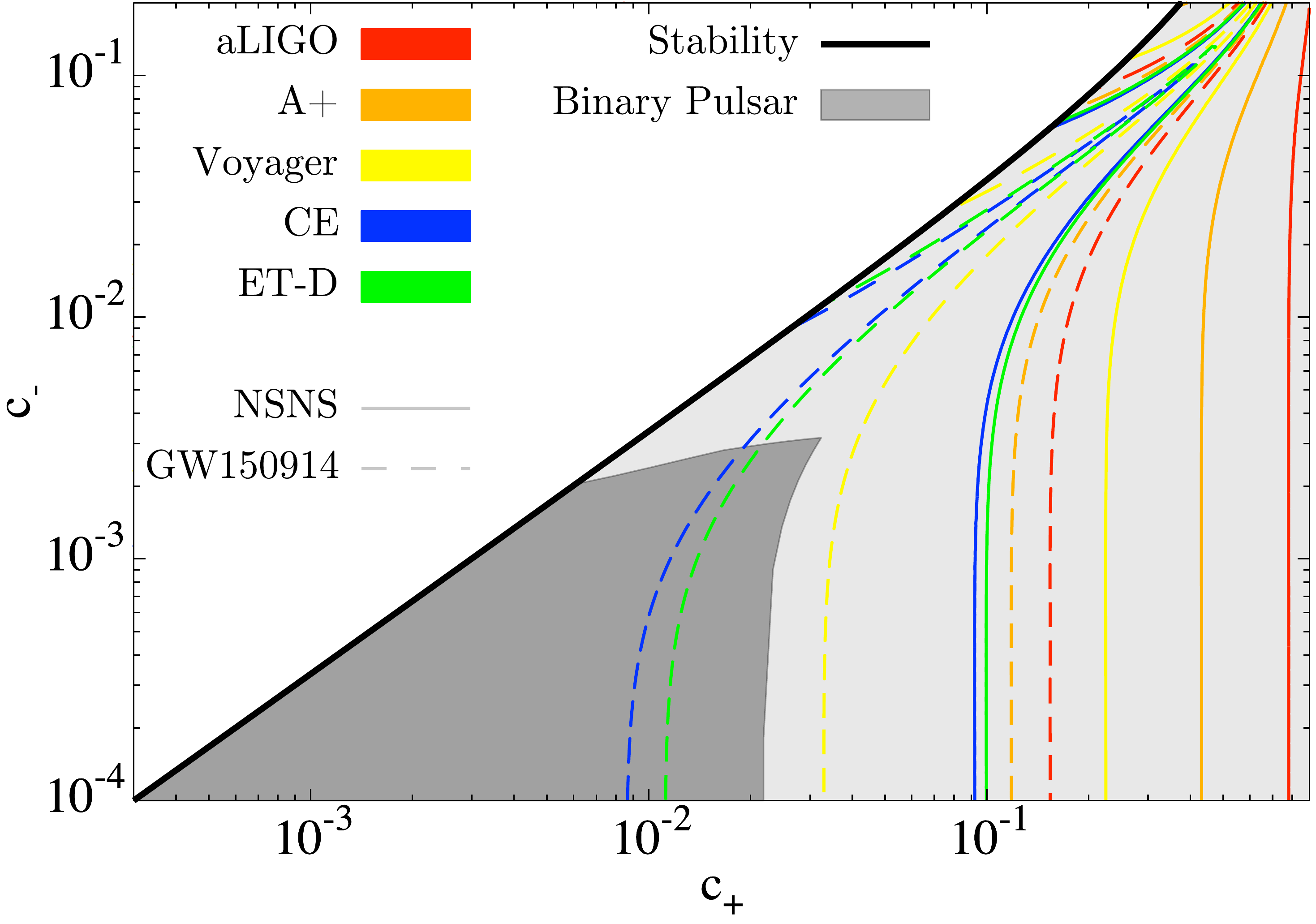}\quad
\includegraphics[width=\columnwidth,clip=true]{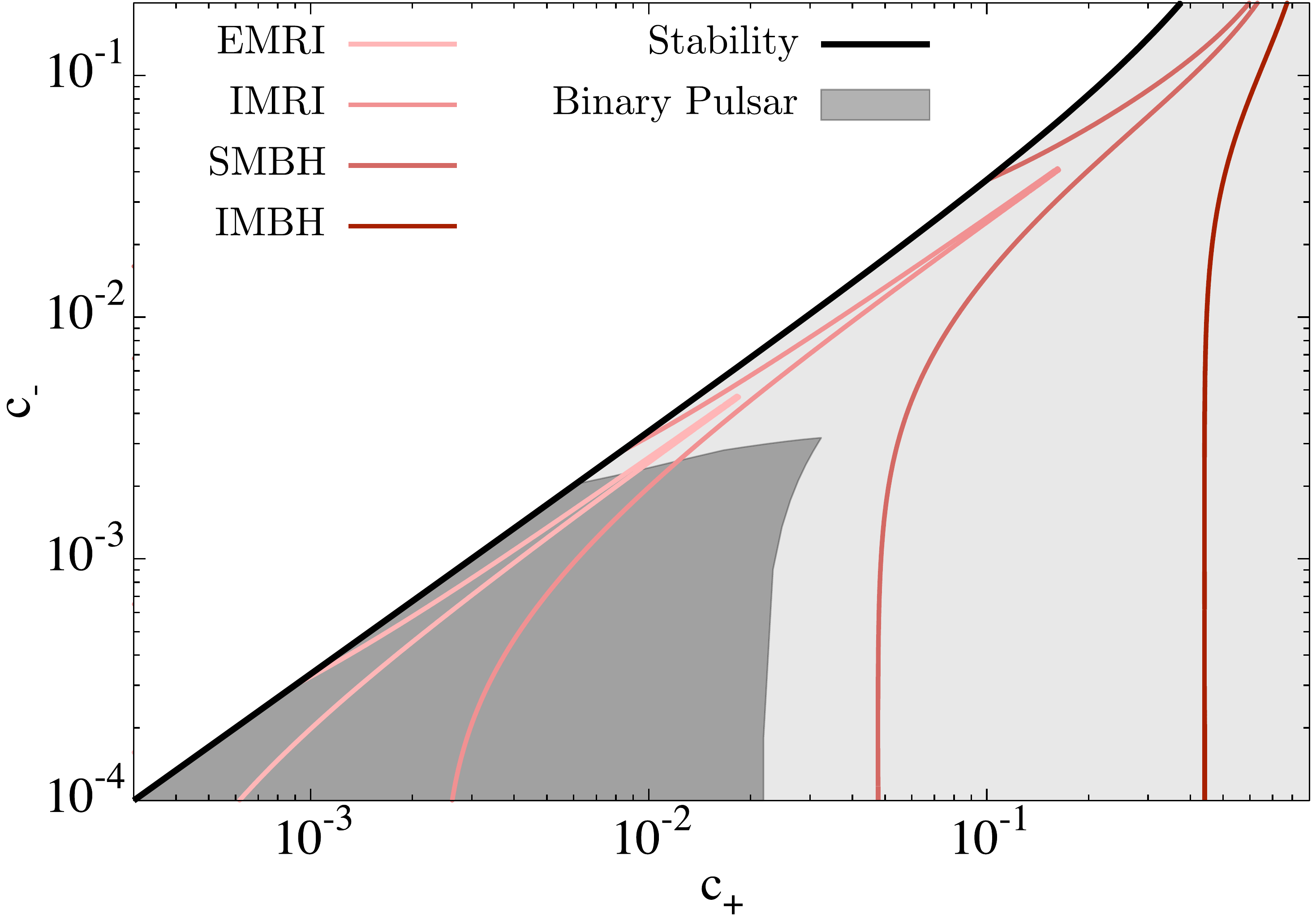}
\caption{\label{fig:EA}(Color Online)
Projected constraint regions placed by ground-based (left) and space-based (right) detectors on the coupling parameters of Einstein-\AE{}ther theory for various systems. The region above the black solid line excludes values of $(c_{+},c_{-})$ that violate certain stability constraints~\cite{Elliott:2005va,Blas:2010hb,Barausse:2011pu}. All values of $(c_{+},c_{-})$ outside of the darker grey region are ruled out by binary pulsar observations~\cite{Yagi:2013qpa,Yagi:2013ava}. The regions above and to the right of the colored lines correspond to values of $(c_{+},c_{-})$ that would be ruled out with future gravitational wave observations using different ground-based instruments.}
\end{center}
\end{figure*}

\subsection{Local Lorentz Symmetry Violation}
In Lorentz violating theories, a vector field is introduced that carries additional energy away from the inspiraling binaries, inducing modifications that enter at -1PN order. These corrections, however, depend on the difference of the compact object sensitivities, which are not known for black holes. We will thus here estimate future constraints on Lorentz violating effects using the sensitivities calculated for neutron stars (the sensitivities of which \emph{have been} calculated) through the next-to-leading order term in the phase, which enters at Newtonian order, as explain in Sec.~\ref{sec:mod-GR}. Once more, the gravitational wave can be modeled as in Sec.~\ref{subsec:waveform-model} with
\begin{align}
\beta =& -\frac{3}{128}\left[\bigg(1-\frac{c_{14}}{2}\bigg)(\scA_{\EA,1}+S \scA_{\EA,2}+S^{2}\scA_{\EA,3})\right]\,,
\nonumber \\
b =& -5\,
\end{align}
where $S\equiv (s_{1}m_{2}+s_{2}m_{1})/m$ and $s_{1,2}$ are the compact object sensitivities, $\scA_{\EA,1}$ can be found in Eq.~(91) of~\cite{Hansen:2014ewa}, and $\scA_{\EA,2/3}$ can be found in Eqs.~(111) and (112) of~\cite{Yagi:2013ava}.
Notice from the mapping that, in this case, the modification depends on more than a single coupling parameter (e.g.~$c_{\pm}$ in the Einstein-\AE{}ther case) and that since these are dimensionless, there is no additional function of the binary's system parameters required in the conversion. We thus expect a relatively simple two-dimensional mapping between $\beta$ and the coupling constants of the theory, as shown in Fig.~\ref{fig:EA}. Observations of some black hole inspirals and mergers with space-based detectors do best at constraining these modifications because they are able to see the merger phase, which breaks a chirp mass-total mass degeneracy in parameter estimation  (see also discussion in~\cite{Cornish:2011ys,Sampson:2013lpa}). However, ground-based detectors become competitive with space-based detectors when one considers binary black hole systems. The ground-based constraints provided by the observation of neutron star inspirals (solid lines in Fig.\ref{fig:EA}) do not include the merger phase\footnote{The merger of binary neutron stars occurs at kHz frequencies where the detectors are less sensitive and where the simple PhenomD waveform model would not be accurate.} in this study, as described in Sec.~\ref{subsec:waveform-model}.

Similarly to Einstein-\AE ther, khronometric gravity introduces modifications to GR at Newtonian order, and thus we expect black hole observations of the merger (with both space- and ground-based detectors) to do better than ground-based observations of neutron star inspirals\footnote{Note that the constraints that we find are roughly one order of magnitude worse that what was found in~\cite{Hansen:2014ewa}. This is due to a difference in waveforms used when performing the Fisher Analysis. In this paper, we use the spin-dependent PhenomD waveform model, as described in Sec.~\ref{sec:mod-GR}--\ref{subsec:waveform-model}, while in~\cite{Hansen:2014ewa} the Taylor F2 model was used.}. The gravitational wave can be modeled as in Sec.~\ref{subsec:waveform-model} with:
\begin{align}
\beta =& -\frac{3}{128}\left[\bigg(1-\frac{\alpha_{\KG}}{2}\bigg)(\scA_{\KG,1}+S \scA_{\KG,2}+S^{2}\scA_{\KG,3})\right]\,,
\nonumber \\
b =& -5
\end{align}
where $\alpha_{\KG}=2\beta_{\KG}$, $\scA_{\KG,1}$ can be found in Eq.~(91) of~\cite{Hansen:2014ewa}, and $\scA_{\KG,2/3}$ can be found in Eqs.~(121) and (122) of~\cite{Yagi:2013ava}. As before, observations of black hole inspirals and mergers do best at constraining these modifications. We see in Fig.~\ref{fig:KG} that constraints with future space-based detectors would be able to greatly shrink the allowed parameter space. Projected constraints with NSNS systems lie outside of the bounds of this plot.
\subsection{Massive Graviton}
A special relativistic modification to the dispersion relation of gravitational waves to include a mass for the graviton introduces a correction in the gravitational wave phase that enters at 1PN order, as discussed in Sec.~\ref{sec:mod-GR}. With the waveform model of Sec.~\ref{subsec:waveform-model}, the ppE mapping is then
\be
\beta = \frac{\pi^{2}\ \mbox{D}_{0}\ \scM_{z}}{\lambda^{2}}\,,
\quad 
b = -3\,
\ee
where D$_{0}$ is given in Eq. 25 of~\cite{Yunes:2016jcc} and $\lambda_{g}$ is the wavelength of the graviton.

\begin{figure}[htb]
\begin{center}
\includegraphics[width=\columnwidth,clip=true]{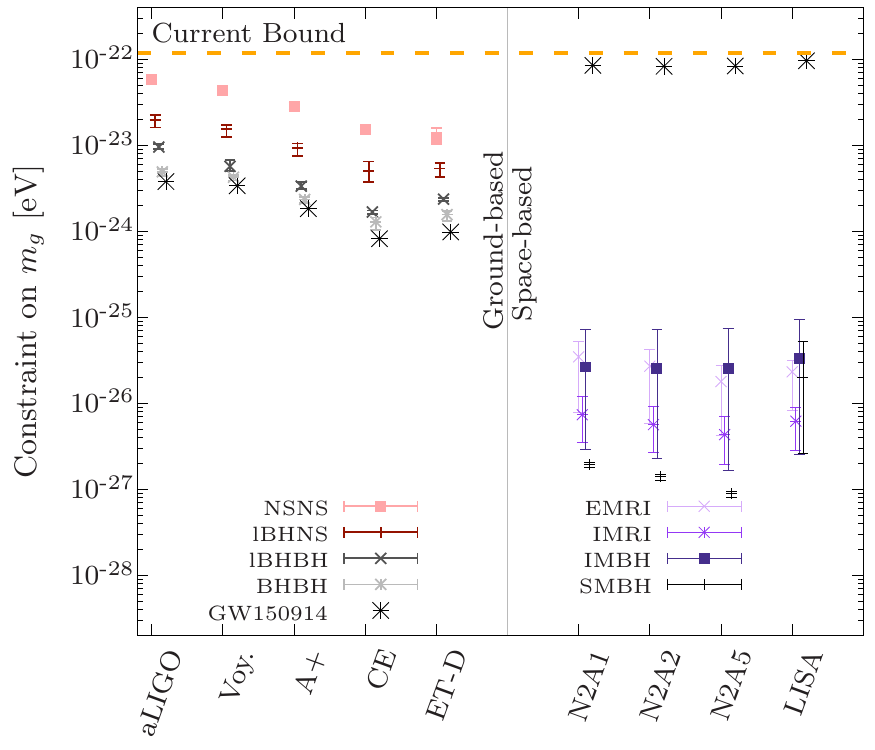}
\caption{\label{fig:mg}(Color Online) Projected constraints on the mass of the graviton as a function of instrument. Notice that SMBH binary systems, which are both the most massive and most distant binary systems considered, can constrain the mass of the graviton up to $\sim$5 orders of magnitude more stringently than current bounds. }
\end{center}
\end{figure}

Notice that in solving for a constraint on $m_{g}$, one must divide by the product of the luminosity distance and the chirp mass. We thus expect that the gravitational waves emitted from the most distant and the most massive systems will lead to the most stringent constraints. This is indeed verified in Fig.~\ref{fig:mg}, where we see the best constraints come from space-based detectors, which can observe supermassive black hole mergers at Gpc distances. These constraints can be as much as 2--3 orders of magnitude better than the best constraints with third-generation ground based detectors. All of these, nonetheless, are as much as 5 orders of magnitude better than current constraints with aLIGO, rapidly approaching the scale at which a mass of the graviton could be comparable to the cosmological constant ($\sim 10^{-31} {\rm{eV}}$). 

\section{Future Directions} \label{sec:conclusions}
We have investigated the constraints we will be able to place on deviations from GR with future space- and ground-based detectors. We found that constraints can improve by more than an order of magnitude as one compares future ground-based instrument observations to current aLIGO bounds. These improvements, however, become much closer to those provided by space-based instruments when considering the future generation detectors. We also quantified the degree to which improvements in different bands of the sensitivity noise leads to improvements in constraints of GR, finding that modest low-frequency improvements can have large effects while high-frequency improvements typically have lesser, but still substantial, effects. We found that this is due to low-frequency improvements that greatly increase the number of effective cycles for certain GR modifications.

The work we have done can be used to extrapolate conclusions about design decisions, but certainly more work could be done to refine the analysis and solidify the conclusions. One example would be to redo the study with a Bayesian analysis instead of a Fisher analysis; we expect this will have a small effect on our conclusions because most of the signals considered have very large signal-to-noise ratio. Another example would be to quantify the systematic errors induced by our approximate waveform modeling in their impact of constraints on deviations of GR; we expect this will also have a small effect for binaries that are widely separated, but the modeling must certainly be improved when considering EMRIs or to include the effects of spin precession. One could also consider the effect of stacking multiple signals on the constraints derived here~\cite{stacking-future}; we expect this to improve the constraints by a factor of roughly $N^{1/2}$ when stacking $N$ signals, but this could affect space- and ground-based instruments differently as they may detect a very different number of sources (since they observe very different populations). A final simple extension would be to consider constraints with multi-wavelength observations (i.e., with both ground- and space-based detectors); given the analysis in~\cite{Barausse:2016eii}, we expect multi-wavelength observations to improve constraints by a factor of a few.  
\begin{acknowledgments}
We thank Rana Adhikari and Matthew Evans for providing noise curve data for future ground-based interferometers, and Neil Cornish for providing noise curve data for proposed LISA. We also thank Emanuele Berti, Antoine Klein, and Kent Yagi for useful discussions. K.C. acknowledges support from the Undergraduate Scholars Program and the Montana Space Grant Consortium Apprenticeship Program.  N.Y. acknowledges support from the NSF CAREER Grant PHY-1250636 and NASA grant NNX16AB98G. 
\end{acknowledgments}
\bibliography{bibliography.bib}
\end{document}